\documentclass[a4,12pt]{article}

\usepackage{mathrsfs,amsbsy,amssymb,latexsym,amsfonts,amsmath}

\catcode`\@=11
\@addtoreset{equation}{section}

\global\arraycolsep=3pt
\newcommand\CL{\mathcal{L}}
\newcommand\CD{\mathscr{D}}
\newcommand\CO{\mathcal{O}}
\newcommand\CS{\mathscr{C}}
\newcommand\bC{\mathbb{C}}
\newcommand\e{\mathrm{e}}
\newcommand\bZ{\mathbb{Z}}
\newcommand\bR{\mathbb{R}}
\newcommand\CU{\mathscr{U}}
\newcommand\Ds{D\hspace{-2.5mm}/}

\begin{document}

\begin{flushright}
{
YITP-06-62\\
OIQP-06-18\\
hep-ph/0612032
}
\end{flushright}
\bigskip

\begin{center}
{\Large 
Future Dependent Initial Conditions from Imaginary Part in Lagrangian}

\hspace{10mm}

{\large
Holger B. Nielsen}\\ 
{\it
Niels Bohr Institute,\\17, Blegdamsvej, DK2100 Copenhagen, Denmark
}
\\
and
\\
{\large
Masao Ninomiya }
{\it
\footnote{Working also at Okayama Institute for Quantum
Physics, Kyoyama-cho 1-9, Okayama City 700-0015, Japan.}\\ 
Yukawa Institute for Theoretical Physics,\\
Kyoto University, Kyoto 606-8502, Japan}
\end{center}

\vfill
\begin{abstract}
We want to unify usual equation of motion laws of nature with ``laws" 
about initial conditions, second law of thermodynamics, cosmology.
By introducing an imaginary part -- of a similar form but different 
parameters as the usual real part -- for the action to be used in the 
Feynmann path way integral we obtain a model determining 
(not only equations of motion but) also the initial conditions, 
for say a quantum field theory.
We set up the formalism for e.g. expectation values, classical approximation
in such a model and show that provided the imaginary part gets unimportant
except in the Big Bang era the model can match the usual theory.
Speculatively requiring that there be place for Dirac strings and 
thus in principle monopoles in the model we can push away  the effects of the
imaginary part to be involved only with particles not yet found.
Most promising for seeing the initial condition determining effects from
the imaginary part is thus the Higgs particle.
We predict that the width of the Higgs particle shall likely turn out
to be (appreciably perhaps) broader than calculated by summing usual 
decay rates.
Higgs machines will be hit by bad luck.
\end{abstract}
\newpage

\section{Introduction}
Usually when we talk about ``theory of everything" as 
superstring theory is hoped to be, 
it is not really meant that the initial state
of the universe is included in the model immediately.
Rather one needs to make extra assumptions -- cosmology, 
second law of thermodynamics\cite{1,2,3,7}, etc. --
about the initial conditions or one simply leaves it for the applicator 
of the theory to somehow himself manage to find out what the initial
conditions are for the experiment he wants to describe with the theory.
It is, however, the intention of the series of articles \cite{slaw,4,5,
6,22} to which 
this article belongs to set up assumptions telling the initial conditions 
in a way that can be called that these initial condition assumptions are 
unified\cite{9} with the part of the theory describing the equations of motion
and the particle content (the usually T.O.E.).
Our unification may though be mainly a bit formal in as far as our main
point is to use in the Feynmann path integral an action which has both
a real $S_R$ and an imaginary part $S_I$.
Usually of course the action is real and the imaginary part $S_I=0$ 
(Total $S=S_R + iS_I$).
We may quickly see that the imaginary part gives a typically hugely
different extra factor in the probability for different paths obeying
equations of motion.
Thus such an imaginary part essentially fix the path obeying equations
of motion which should almost certainly be the realized one.
In this way we can claim that to a good approximation an imaginary part
of the action will choose/settle the initial conditions.

In the present article it is not the point to settle on any choice of 
the in usual sense ``theory of everything".
Rather we shall present our idea of introducing an imaginary part in the
Lagrangian and thereby also in the action as a modification that can be
made on any theory as represented by the real action $S$.

We have already published a few articles on essentially a classical 
formulation of the present model.
We sought in these articles to be a little more general by simply defining
a probability weight called $P{(\mathrm{path})}$ defined for all possible
paths.
In classical theory it is really only the paths which obey the classical 
equations of motion for which we need to define $P$.
We already in the earlier articles suggested that this probability 
$P{(\mathrm{path})}$ for a certain track, path, to be the one realized in
nature should be given as the exponential of an expression depending on
the track, path, of the form of a space-time integral over a locally
defined quantity $\CL_I$ depending on the fields in the development, path.
Really this quantity $\CL_I$ (really $-\CL_I$) comes into determining the 
probability as if it were the imaginary part of the Lagrangian density.

A major point of the present article is to set up the quantum formulation
of our already published model, now really settling on taking the suggestive
idea of just making the action complex, but with a priori a different set
of coupling constants and $m^2$ for real and imaginary part separately.

A genuine problem with our kind of model is that very likely it predicts
that special simple configurations leading to big probability may be 
arranged at a priori any time.
That is to say, with our type of model it needs an explanation that one 
in particle almost never see any great arrangements being organized to occur
later on.
Really such arrangements might seem to us to be something like a hand of God,
but they seem very seldom.
Thus at first it looks that our type of model is already falsified by the
non-appearance of arrangements. Really such a problem is almost obviously 
expected to occur in a model that like ours {\em does not a priori make any 
time 
reversal asymmetric assumption at the fundamental level}. Unless in the 
Hartle-Hawking no boundary postulate \cite{9} we add some timereversal 
asymmetry spontaneously other otherwise that theory will be up to similar 
problems \cite{8,23}.

A model-language  describing how final states can be imposed by a density 
matrix $\rho_f$ is put forward by Hartle and Gell-Mann \cite{hg}.   

In the present work we hope for that a certain 
moment in the `middle of times' will turn out to become dominant w.r.t.
fixing the special solution selected as the realized one, and that this time 
can then be interpreted as a close to Big Bang time ( there may not really be 
a true big bang but just an inflation era coming out of a deflation era 
continuously). Then since we live in the time after this decisive Big Bang 
simulating era there is for us a time reversal asymmetry, nevertheless it
is a problem that like ours is even timetranslational invariant w.r.t. the 
law that finally settle the `initial conditions' to explain that there 
are not more prearranged events than one seemingly see.

However, we believe to have found some explanations able to 
suppress so many of these prearrangements that our model can be 
made compatible with present experience of essentially no prearrangements.

For really avoiding it we shall assume consistency of Dirac strings, but let
us postpone that discussion to section 13 below.

Our model is really inspired from the considerations of time machines\cite{tm} 
and the 
troubles of needs for prearrangements in order to avoid the so called grand 
mother paradoxes, meaning the inconsistencies occurring when one seeks to go 
back in time and changes the events there.

We shall present the work by making {\em two} attempts to assumption about how 
to 
interpret the Feynmann path integrals with the imaginary part of the action 
non-zero. In the first part of the paper we start out from letting the average 
of a dynamical variable ${\cal O}$  be given by equation (2.9) below, but 
that this is a priori not so good is seen by it not being (safely) real 
even if the 
dynamical 
variable ${\cal O}$ is real. Therefore in section 7 we restart the discussion
so now from the side of the interpretation of the Feynmann path way integrals 
in our model.

{\bf First trial of interpretation}

In the next section 2, we shall put forward the basic formula for expectation
values with our complex action model and the philosophy that this model
even deliver the initial condition, or better the solution of equation of
motion to be the one realized.

In section 3 we review our earlier reasons for that future should have only
little influence on what happens.

In section 4 we then shall argue for some approximate treatment of the 
functional integral in the late times $t$, the future.

In section 5 we shall make use of the approximation of the future to obtain
the usual quantum mechanics expressions at least in the case where our
imaginary part $S_I$ of the action can be ignored.
(It should be stressed that we actually have used already a philosophy 
based on this $S_I$ being non zero, so it is not fully zero.)

In section 6 it turns out that we -- perhaps not completely convincing
though -- can make the effect be that we return to probability in practical
scattering experiments say get conserved.

{\bf Second trial of interpretation:}

In section 7 we restart the discussion of making the interpretation 
formula for the Feynmann path way integral, which after some talk takes 
the way of using the classical approximation weighted with the exponential 
of minus 2 times the imaginary part of the action. 
In a subsection 7.5 we formally connect
our model to our  earlier one based on the probability weight 
$P{(\mathrm{path})}$.

In section 8 we develop a rather general formula for the correlated 
probability for a series of dynamical quantities or operators ${\cal O}_i$
at different moments of time take values inside small ranges specified.

In section 9 we go a bit further in making the expressions like the ones 
one uses in practice in usual theories. Most importantly we again consider 
how to approximate the future when the effects of the imaginary part of the 
action is very small.

In section 10 we put the  simplest example of the more general formula,
namely a formula for the probability of just one operator at one time 
being in a given range. ( This question would be impossible to predict 
even in principle in other theories, but we in principle can, but in practice 
not usually). But the resulting formula has what we call ``squared form'' in 
the sense that the projector comes in twice as a factor in it. The finding 
of a reduction to an unsquared form is left to section 12, while we 
in section 11 then give an example of application of very interesting 
physical significance. In fact section 11 predicts a broadening of the 
width of the Higgs particle due to the imaginary action.

In section 12 we then bring about a connection between the postulated 
interpretation formulas for probabilities put forward in part I and part II.
In fact we find that they coincide under rather suggestive  assumptions.

In section 13 we bring the promised argument for removing the effects
of the imaginary action $S_I$ from the domain of older accelerators, since
otherwise our model would have been falsified. The argument is based on
assuming monopoles.

In section 14 we conclude and give a bit of outlook.
\vspace{.5 cm}

\noindent
{\huge \bf Part I, First Trial of Interpretation}

\section{Philosophy and formula}
Our basic modification of introducing an imaginary part in the actions leads 
to that integrand $\e^{iS}$ or $\e^{\frac{i}{\hbar} S}$ of the Feynmann path 
way integrand
$\int \e^{iS} \CD \phi$ or $\int \e^{\frac{i}{\hbar}S} \CD \phi$
(if the Planck constant is written explicitly) varies a lot in magnitude,
and not only in phase as usual.
This effect is likely to make some regions in the space of paths 
-- or we could restrict to the space paths with $\delta S=0$, i.e. the space
classical solutions -- 
get a very much bigger weight in the integral than others.
Actually it can likely happen that only a very narrow range of paths or
better solutions (= paths obeying $\delta S=0$) will quite dominate the 
integral
\begin{eqnarray}\label{1}
\int \e^{iS} \CD \phi .
\end{eqnarray}
 
That should naturally be taken to mean that the presumed narrow range of
dominating paths represent the paths being actually realized in nature.
It is in this way that we hope our model to essentially predict the initial
state for the realized solutions.
It is important to have in mind that such an effect of the imaginary action
$S_I$ of selecting narrow bunches of solutions can make the boundary 
conditions at an initial and a final time for a period to be studied say,
superfluous.
A bit optimistically we might imagine that the imaginary part of action 
makes the functional integral converge even without boundary condition
specifications.
Note that being allowed to throw boundary conditions away
-- having them replaced by effects of $S_I$ --
is a great/nice simplification.
We consider this achievement as an aesthetically very nice feature of our
model!
Supposing that this works to deliver a meaningful Feynmann-path integral
(\ref{1}) even without boundary conditions this way we must now decide how
one is supposed to extract information now in principle for the true
expectation value as it should occur even without further input.
Note here that we are -- but only in principle --
proposing an exceedingly ambitious model compared to usual quantum field
theories:

We want to predict expectation values without any further input than the
mere complex action!
This of course corresponds to that our level of ambition is to in addition
to the usual time-development laws of nature \underline{also} predict the
initial conditions, i.e. what really happens!

To write down the formula for some physical quantity let us first exercise
by a quantity $\CO (\varphi |_t)$ which is a function of the fields $\varphi |_t$
restricted to some time $t$, where $\varphi$ is a general symbol for all the
fields in the model.

If we for instance use the Standard Model as the starting model, then
providing it with an imaginary part of the Lagrangian density, then the 
symbol $\varphi (x) \,(x\in \grave{R}^4)$ is really a set
\begin{eqnarray}\label{2}
\varphi = (A^a_\mu, \psi^b, \,H)
\end{eqnarray}
where the indices on the Fermion fields runs through the combination of
flavor and color and/or $W$-spin components, while the index on the 
gauge fields run through the 12 gauge fields
-- 8 gluon color combination plus $(B_\mu)$ the $U (1)$-component
and 3 $W$'s --.
Finally $H$ is the two complex component Higgs field.

The quantity $\CO(\varphi |_t)$ can of course be considered a functional of
the whole field development $\CO(\varphi)$ also, i.e. it could be consider
a functional of  the path of one wants.

The simplest proposal for what the average quantity $\CO(\varphi)$ 
would be
\begin{eqnarray}\label{3}
\langle \CO(\varphi) \rangle = \frac{\int \e^{iS[\varphi]}\CO(\varphi)\CD\varphi}
{\int \e^{iS[\varphi]}\CD \varphi}.
\end{eqnarray}
This would mean that we have a ``sort of probability" given by
\begin{eqnarray}\label{4}
\mathrm{``Probability\, of}\, \CO \,\mathrm{being} \,\CO_0"
=\frac{\int \delta(  \CO(\varphi)-\CO_0) \e^{iS[\varphi]}\CD\varphi}
{\int \e^{iS[\varphi]}\CD\varphi}
\end{eqnarray}
Now, however, we must admit that conceiving of this expression as  a 
probability is upset by the severe \underline{problem} that it will 
typically be a complex number.
There is no guarantee that it is positive or zero.

Thus a priori one would say that this simple expression for the probability
density is quite untenable.

Nevertheless it is our intention to claim that we should 
-- and that is then part of our model -- 
use the simple expression (\ref{4}) and the corresponding (\ref{3}) and 
the expression to be given below for more general operators $\CO$ 
corresponding also to (\ref{3}) and (\ref{4}).

First let us again stress that it is our a priori philosophy that somehow
the imaginary part $S_I$ managed to fix both a state in future and 
in past.
Thereby asking the average of some quantity $\CO$ becomes much like 
in an already finished double slit experiment (Bohr-Einstein) in which
a particle already have been measured on the photographic plate
(presumable on an interference line) what \underline{were} the average
position of the particle when it past the double slit screen.
Really asking such a question concerning a quantity $\CO$ that were not 
measured and could not have been measured without having disturbed the
outcome of something later is one of the forbidden questions in quantum
mechanics.
Indeed it is by asking this sort of questions which are not answerable by
measurement that Einstein can find ammunition against quantum mechanics.
In other words our proposal (\ref{4}) for ``probability distribution" is 
a priori -- with our present philosophy of a future essentially determined
by $S_I$ -- an answer to a quantum mechanically forbidden question.
Niels Bohr would say we should not ask it.

In that light it may of course not be so serious that our formula gives
a rather stupid or crazy answer, a complex probability!

But now we have the problem of justifying that if we made a true measurement
the answer would turn out to give positive (or zero) probability.

Let us take as the important feature of a measurement of some quantity 
$\CO$  that there is an apparatus which makes a lot of degrees of freedom,
$\xi$ say (really macroscopic systems) develop in a way depending on value 
of $\CO$.
Such an amplification of the effect of the actual value of $\CO$ is
characteristic for a measurement.
Unless somehow there are special reasons for that $S_I$ be insensitive to $\xi$
(as we shall actually later seek to show but do not assume to be the case)
we expect that $S_I$ typically will depend on the macroscopically many
d.o.f. $\xi$ being influenced by $\CO$-value measured.
Now we argue like this:  Since there is a huge (macroscopical) number of
variables $\xi$
depending on the value of $\CO$ ``measured", the imaginary part $S_I$ of
the action is likely to depend very strongly on this measured value -- very
rapidly varying.

We here think of $S_I$ as the integral over the imaginary part of the
Lagrangian $L_I$ over all times $t \in ] -\infty, \infty [ $
\begin{eqnarray}
\left(
S_I = \int_{-\infty}^{\infty} L_I \mathrm{dt} 
\right)
\end{eqnarray}

Because of the great complications in an actual measuring apparatus, let
alone the further developments  depending the measured value, publications and 
so on, the imaginary action $S_I$ can easily be a very complicated 
function of the measured $\CO$ value.
Even if $S_I$ as function of the measured $\CO$ value should in principle
be continuous it may in practice vary so much up and down
-- caused by accidents influenced by the broadcasted measuring value --
that very likely the smallest value of  $S_I$  occurs for a seemingly
accidental value of the measured $\CO$.
If the $S_I$-variation with the ``measured $\CO$" is indeed very strong
so that the $S_I$ variations are big the exponential weight $\e^{-S_I}$
contained in (\ref{3}) and (\ref{4}) will have a completely dominant
value for only one measured $\CO$-value.

In this way our model has the integral in the numerator of (\ref{4}) be
much bigger for one single value of $\CO_0$.
If so, then the ratio (\ref{4}) is actually $\propto 1$ for this 
$\CO_0$-value and negligible for all other  $\CO_0$-values.
This means that our model much like usual measurement theory
(in Copenhagen interpretation) predicts that crudely only one value of 
a measured quantity is realized.
In principle it is even so that, bearing a very special situation,
the result of the measurement is calculable by essentially minimizing
the imaginary action $S_I$.
In practice, however, such calculation will only be doable in extremely 
rare cases.
(If we impress a special result by threatening with a Higgs-producing
machine).

We postpone the argumentation for that the probability distribution to be
obtained in practice shall be the one of usual quantum mechanical
measurement theory partly to the later sections and partly to a 
subsequent paper.

At the end of this section let us extend slightly our formula (\ref{3})
and thereby also (\ref{4}), to the case where the quantity $\CO$ 
corresponds in usual quantum mechanics to an operator that do not commute
with the fields $\varphi$.

An operator corresponding to a quantity measurable at a moment of time $t$
will in general in the quantum field theory considered be given by a 
matrix with a columns and rows in correspondence with field functions
$\varphi |_t$ restricted to the time $t$.
I.e. $\CO$ is given by a ``matrix"
\begin{eqnarray}\label{5}
\left(\varphi' |_t \big| \CO \big| \varphi |_t\right)
= 
\underbrace{\CO (\varphi' |_t, \,\varphi |_t)}
.
\end{eqnarray}

What should be the formulas replacing (\ref{3}) and (\ref{4}) in this
more general case?

Well, our main starting point were that we assumed our imaginary part $S_I$ 
to (essentially) fix both a further $| B \rangle$ and a past state 
$| A \rangle$.
A natural notation to introduce is in fact -- for the past --
\begin{eqnarray}\label{6}
\langle \varphi|_t | A \rangle
= A[\varphi|_t ]
= \int_{\mathrm{ending\, at}\, \varphi|_t
} \e^{iS_{-\infty \,\mathrm{to}\,t}}\CD \varphi
\end{eqnarray}
and analogously
\begin{eqnarray}\label{7}
\langle B|\varphi|_t  \rangle
= \langle \varphi|_t | B \rangle^{*}
= B[\varphi|_t]^{*}
= \int_{\mathrm{beginning\, at}\, \varphi|_t
} \e^{iS_{t\,\mathrm{to}\,+\infty }}\CD \varphi
\end{eqnarray}
 
In this notation our previous formulas (\ref{3}) and (\ref{4}) are for
the in $\varphi|_t$ diagonal operators $\CO(\varphi|_t)$ become
\begin{eqnarray}\label{8}
\langle \CO \rangle &=&
\frac{\int \e^{iS}\CO(\varphi|_t)\CD\varphi}
{\int \e^{iS}\CD\varphi}
\nonumber\\
&=& \frac{\oint _{\varphi|_t}\langle B|\varphi|_t  \rangle \CO(\varphi|_t)
\langle \varphi|_t | A \rangle}
{\oint _{\varphi|_t}\langle B|\varphi|_t  \rangle \langle \varphi|_t | A \rangle}
\nonumber\\
&=& \frac{\langle B|\CO|A \rangle}{\langle B|A \rangle}
\end{eqnarray}
and (\ref{4}) becomes
\begin{eqnarray}\label{9}
\mathrm{``Probability\, of}\, \CO \,\mathrm{being} \,\CO_0"
=\frac{\langle B|\delta (\CO-\CO_0)|A \rangle}
{\langle B|A \rangle}
\end{eqnarray}
Now really we want to suggest that formula (\ref{8}) and (\ref{9}) can
also be used for operators that are not simply functions of the 
fields $\varphi|_t$ at
time $t$,  used in the functional integral.

In order to justify that extension of our interpretation formulas 
we want to remark:
\begin{enumerate}
  \item Provided a Hermitean operator $\CO$ has either $|A\rangle$ or
$|B\rangle$ as eigenstate then the eigenvalue $\CO'$ in question can of course
be extracted as
\begin{eqnarray}\label{10}
\CO'=
\frac{\langle B|\CO|A \rangle}
{\langle B|A \rangle}
\end{eqnarray}
 \item One can quite generally 
-- by Fourier transformations at every step in a time lattice --
rewrite a functional integral of the Feynmann path way integral form
from some set of variables $\varphi$ to a conjugate set:
\begin{eqnarray}\label{11}
\int \e^{iS}\CD \varphi &\stackrel{\mathrm{latticitation}}{=}&
\int \prod_{t\in \{ t \mathrm{-lattice}\}} \CD^{(3)} \varphi |_t
\e^{i\sum_{t\in \{\mathrm{lattice}\}} L_{\mathrm{discr}}
\left( \varphi|_t , \frac{\varphi|_{t+\Delta t}-\varphi|_t}{\Delta t}
\right) \Delta t}
\nonumber\\
&=& \int \prod_{t\in \{ t \mathrm{-lattice}\}} \CU
(\varphi|_{t+\Delta t}, \varphi|_t) \CD^{(3)}\varphi|_t
\end{eqnarray}
\end{enumerate}
where
\begin{eqnarray}\label{12}
\CU (\varphi|_{t+\Delta t}, \varphi|_t)
= \e^{i L \left( \varphi|_t, \frac{\varphi|_{t+\Delta t}- \varphi|_t}
{\Delta t}\right)},
\end{eqnarray}
can be rewritten into $\hat{\CU}(\Pi |_{t+\Delta t}, \Pi |_t )$
matrices obtained from the $\CU(\varphi |_{t+\Delta t}, \varphi |_t )$
by Fourier functional transformations
\begin{eqnarray}\label{13}
\hat{\CU}(\Pi |_{t+\Delta t}, \Pi |_t ) \stackrel{\mathrm{def}}{=}
\int \CD^{(3)} \varphi |_{t+\Delta t}
\,\e^{+i \varphi |_{t+\Delta t} \Pi |_{t+\Delta t}}
\CU(\varphi |_{t+\Delta t}, \varphi |_t)
\e^{-i \varphi |_t \Pi |_t} \CD^{(3)} \varphi |_t
~~~
\end{eqnarray}
Now of course for long chains of $\hat{\CU}$-matrices 
(ignoring end problems) you have
\begin{eqnarray}\label{14}
\int \prod_{t\in \{ t \mathrm{-lattice}\}} \CU
(\varphi|_{t+\Delta t}, \varphi|_t) \CD^{(3)}\varphi|_t
\nonumber\\
\stackrel{\mathrm{except\, for\, end\, problems}}{=}
\int \prod_{t\in \{ t \mathrm{-lattice}\}} \hat{\CU}
(\Pi |_{t+\Delta t}, \Pi |_t) \CD^{(3)}\Pi |_t
\end{eqnarray}
Supposedly you can put the right hand side into a form
\begin{eqnarray}\label{15}
\int \e^{i S^{(\mathrm{in}\,\Pi)}[\Pi, \,\Delta \Pi \mathrm{-defferences}]}
\CD \Pi
\end{eqnarray}

Now you may argue with the same intuitive suggestion for getting
\begin{eqnarray}\label{16}
\CO (\Pi |_t)=
\frac{\int \e^{i S^{(\mathrm{in}\,\Pi)}}\CO(\Pi |_t) \CD \Pi
}
{\int \e^{i S^{(\mathrm{in}\,\Pi)}}\CD \Pi}
\end{eqnarray}
as we did for (\ref{3}).
By thinking of doing the just presented Fourier transformation partly we
might argue for a similar average formula for any operator
\begin{eqnarray}\label{17}
\langle \CO (\varphi |_t, \,\Pi\_t)\rangle &=&
\frac{\int \e^{i S} \CO(\varphi |_t, \Pi |_t) \CD \varphi}
{\int \e^{i S} \CD \varphi}
\nonumber\\
&=& 
\frac{\langle B_t | \CO(\varphi |_t, \Pi |_t)| A_t \rangle }
{\langle B_t|A_t \rangle}.
\end{eqnarray}
Really this proposal looks very bad because of several lacks of good
correspondence with usual quantum mechanics a priori:
\begin{enumerate}
  \item[a)] Obviously $|A_t\rangle$ is here (a sort of) wave function of the 
universe at time $t$, but our probability density (\ref{4}) or
\begin{eqnarray}\label{18}
\mathrm{``Probability\, for}\, \CO \,\mathrm{being} \,\CO_0"
=\frac{\langle B_t | \delta (\CO - \CO_0) | A_t \rangle }
{\langle B_t|A_t \rangle}
\end{eqnarray}
is not quadratic in $|A_t \rangle$ as we expect from the usual corresponding
formula 
\begin{eqnarray}\label{19}
\mathrm{``Probability\, for}\, \CO \,\mathrm{being} \,\CO_0 \,\mathrm{usual}"
=\frac{\langle A_t | \delta (\CO - \CO_0) | A_t \rangle }
{\langle A_t|A_t \rangle}.
\end{eqnarray}
  \item[b)] As already stated the ``probability density" (\ref{18}) is
even usual complex and  needs the above measurement special case to become
just positive. 
\end{enumerate}

We shall below argue for an approximate treatment of the future part 
$|B_t \rangle$ of the integral thereby achieving indeed a 
{\it rewriting into an expression which is of the form with $|A_t \rangle$
coming squared}.
Indeed we shall rewrite (\ref{18}) into (\ref{19}) below.

\subsection{
Justification of philosophy from semiclassical approximation}
In semiclassical approximation one simply evaluates different contributions 
to the functional integral $(1)$ by seeking the different extrema
for $\e^{iS}$ or equivalent $S=S_R + iS_I$.
Around such an extemum it is extremely well known that one can approximate
$S$ by the Taylor expansion up to second order
\begin{eqnarray}\label{triangle1}
S &=& S(\mathrm{extremum})
 + \frac{1}{2}\int \frac{\partial^2 S
}{\partial \varphi_{1} (x_1) \partial \varphi_{2}(x_2)}
\nonumber\\&&
\cdot
\left( \varphi_{1} (x_1) - \varphi_{1}^{\mathrm{extr}} (x_1) \right)
\left( \varphi_{2} (x_2) - \varphi_{2}^{\mathrm{extr}} (x_2) \right)
+ \cdots ~~~
\mathrm{d}^{4}x_1\mathrm{d}^{4}x_2
\end{eqnarray}
where then the linear terms
\begin{eqnarray}\label{triangle2}
\int\frac{\partial S}{\partial \varphi_1(x_1)}
\left(\varphi_1(x_1)- \varphi_{1}^{\mathrm{extr}}(x_1)
\right)\mathrm{d}^{4}(x_1)
\end{eqnarray}
vanish because of the extremiticity condition.
Here $ \varphi_{1}^{\mathrm{extr}} (x_1)$ and
$\varphi_{2}^{\mathrm{extr}} (x_2)$ denote the fields at the extremum
field configuration development.
Such an extremum as is well known corresponds to a solution to
\begin{eqnarray}\label{triangle3}
\delta S = 0
\end{eqnarray}
i.e. solving the variational principle leading to classical equations of
motion.

The main term in the exponent $iS$(extremum) is in the usual real action
case purely imaginary and thus only gives rise to a phase factor so that
in this approximation the contribution has the same size for
all the classical solutions, provided they can go on for real field
configurations.
With our $S_I$ included, however, we tend to get even to the approximation
of the first term in the Taylor expansion (\ref{triangle1}) a real term
$-S_I$ into the exponent and thus the order of magnitude for one classical
solution compared to another can easily become tremendous
\begin{eqnarray}
| \e^{iS (\mathrm{extremum})} | = \e^{-S_{I} (\mathrm{extremum})}.
\end{eqnarray}
It is our philosophy that only relatively very few classical solution
have $\e^{-S_{I} (\mathrm{extremum})}$'s dominating violently the rest.
In this sense we expect and assumed that such one or a very few
classical solutions could be considered the only one realized.
With very big size of $S_I$
-- and that can easily come about for a couple of reasons --
it gets relatively only exceedingly few classical solutions that are
competitive in the sense that for most classical solutions
(of (\ref{triangle3})) you have exceedingly small $\e^{-S_I}$ compared
to the few dominant ones.
As the reasons for $S_I$ being big when it is not forbidden by gauge
invariance and the condition that Dirac strings shall be  unobservable we 
can give:
\begin{enumerate}
  \item There is in analogy to the $S_R$-term a $\frac{1}{\hbar}$-factor
in front of $S_I$. 
For practical purposes we know that we shall consider the Planck constant
$\hbar$ to be very small.
  \item We could easily get Avogadros number come in as a factor in
the $S_I$ because it would get such a factor a priori since there are 
typically in the world of macroscopic bodies of that order magnitude molecules.
\end{enumerate}

\section{Approximate treatment of future part of functional integral
(treatment of $|B_t \rangle$)}

In our earlier works\cite{6}
-- in which we mainly worked in the classical approximation --
we presented some arguments that in the era which have been going on since
short time of after some effective (or real) Big Bang the imaginary
Lagrangian or action $L_I$ or $S_I$ effectively became very trivial.
That should mean that under the times starting after some early Big Bang
and extending into the future we could approximately take $L_I$ and
the part of $S_I$ coming from this era as independent of what are the
practical possibilities for what can go on.
Thus we should in this present era supposed to extend into even the 
infinite future be allowed to ignore in first approximation the imaginary 
parts $L_I$ or $S_I$.

The reasons, which we presented for that were that this present era
including supposedly all future is dominated by two types of particles:
\begin{enumerate}
  \item Massless particles (really the entropy of the universe is today
dominated by the massless microwave back ground radiation of photons).
  \item Non-relativistic particles carrying practically conserved
quantum numbers (the nucleons and the electrons are characterized by
their charges and baryon or lepton number so as to make their decays 
into lighter particles impossible).
\end{enumerate}

The argument then went that we could write the action
-- actually both real $S_R$ and imaginary $S_I$ --
for these particles, treated as particles, as a sum having each giving a 
contribution
proportional to the eigentimes for them: 
\begin{eqnarray}\label{3.2}
S_R, \,S_I = \sum_{\mathrm{particles}\, P} K_{P \{{ R\atop I} \} } \cdot \tau_P .
\end{eqnarray}

That is to say that each of the particles contribute to $S_I$ say a 
contribution proportional to the eigentime
\begin{eqnarray}
S_{I \,\mathrm{from}\, P} \propto \tau_P .
\end{eqnarray}

Now for massless particles any step in eigentime
\begin{eqnarray}
\Delta \tau_P =0~~ \mathrm{(for\, massless)}
\end{eqnarray}
and for nonrelativistic ($\simeq$ slow) particles, such a step is
\begin{eqnarray}
\Delta \tau_P = \Delta t
\end{eqnarray}
equal to the usual time.
Since the number of the conserved quantum numbers protected
particles are all the time the same the whole contribution to the $S_I$
from the present era becomes very trivial:

Zero from the massless, and just a constant integrated over coordinate
time for the conserved particles.

In addition there are terms  from interactions contributing a priori to
say $S_I$ also.
Since, however, in the era since a little after Big Bang the density of
particles were low in fundamental units presumably also the interaction
contributions would be much suppressed in this after Big Bang era.

So all together we estimate that it is only the very early Big Bang times
that will dominate $S_I$.
Thus the solution to the equations of motion being in a model with an
imaginary action $S_I$ selected to be the realized one will mainly depend 
on what happened in that solution in the early Big Bang era.
This means that it will be in our era as if it were the initial state that
were a rather special one determined by having an especially small
contribution to $S_I$ from Big Bang times.
This would mean a rather well determined starting state roughly which 
interpreted as a macrostate would be one with low entropy.
That is at least a good beginning for obtaining the second law of
thermodynamics, since then there are supposedly no strong effects of 
$S_I$ any more to enforce the universe to go to any special macrostate.
Rather it will go into bigger and bigger macrostates meaning that they
have higher and higher entropy.

Although we have now argued for approximately seeing no effects of $S_I$
in the era after Big Bang implying that our model should have no effects
in this era, this is however, presumably not
being quite sufficiently accurate.

We shall, however, below in section 9 invent or find arguments that will 
allow us to get completely rid of the $L_I$ or $S_I$ from the in the
Standard Model already found particles.
Only for the Higgs involving processes our arguments in section 9
based on gauge symmetry and the assumption of unobservability of Dirac
strings associated with monopoles
will not quite function.
Thus we still expect that an $S_I$-contribution pops up with Higgs-particles.
But since Higgs-particles are so far not well studied such an effect of
$S_I$ might well have been overlooked so far.

\section{Treatment of $|B_t \rangle$ or
Treatment of the future factor in the functional integral}

In equation (\ref{7}) above we defined what one could
call ``the future part" of the functional integral relative to the time
$t$.
It should however be kept in mind that it is a part in the sense that the 
full integral is a contraction (a sort of product) of the past part and 
this future part,
\begin{eqnarray}\label{3.1}
\int \e^{iS}\CD \varphi = \langle B_t|A_t \rangle .
\end{eqnarray}
Now we must remember that according to the second law of thermodynamics
the state of the universe if at all obtainable (calculable) should be so
by considering the development in the past having lead to it.
The future, however, should be rather shaped after what happened earlier.
This suggests that we should mainly have the possibility to guess or know
$|A_t \rangle$ but determined from the fundamental Lagrangian as our model
suggests.
Really in order not to disagree drastically with the second law of 
thermodynamics the future should be shaped from the past and reflect the
latter.
However, there should not be -- at least not much --
adjustment of the happenings at say time $t$ in order to arrange something
special simple happening in future.
This means in or formalism that the by the $S_I$ future contributions
determined $|B_t \rangle$ should according to second law better disappear
quite from our formula for predicting probabilities for operator values,
i.e. from (\ref{4}) or more generally (\ref{18}).

Now, however, as we argued in foregoing section -- section 3 --
reviewing previous articles working in the classical approximation
it should be the state of a solution to the equations of motion in the
early Big Bang time that dominates the selection of such a solution to
be the realized one.
The future on the other hand has only a small effect, if any, on
choosing the true or realized solution.
With the arguments to be given in section 9 we argue for the effects of
$S_I$ being even smaller in the future.
Nevertheless we have if we talk exactly also effects of $S_I$ even in
the future.
Otherwise the hypothesis that the integral (\ref{7}) defining $| B_t \rangle$
would be senseless since the $\e^{- S_I}$-weighting is needed to suppress
the integrand $\e^{-S_I}$ enough to make hope of a sensible practical
convergence.

However, we have in section 3 and will in section 9 argue for that 
$S_I$ varies much less in the future than in Big Bang era.

It is now the purpose of the present section to use this only weak $S_I$
variation with the fields in the future to argue for an approximation
in density matrix terminology for the future part $| B_t \rangle$ of 
the functional integral.

Let us indeed perform the following considerations for estimating the
crude treatment of $| B_t \rangle$  which we shall use:
\begin{enumerate}
  \item[a)] Since $S_I$ has in practice only small non-trivial 
contributions in the future it is needed to involve contributions in 
the integral
\begin{eqnarray}\label{4.1}
S_{It' \,\mathrm{to}\,+\infty}
=\int_{t'}^{\infty} \mathrm{dt} \, \int \mathrm{d}\vec{x} \, L_I
\end{eqnarray}
from very large $t\ge t'$.
  \item[b)] At these enormous $t$ regions then at the end we get finally
a rather restricted range of solutions.
-- we can think of classical solutions here, if we like --
  \item[c)] Now the solutions from the enormously late times under a) 
have to be developed backward in time to the time $t'$ say to deliver
the state $| B_{t'} \rangle$ 
(really we first get $\langle  B_{t'}|\phi \rangle$ from equation 
(\ref{7})).
  \item[d)] Now we make the assumption that the system/world is 
sufficiently ``ergodic" and the large times so large and so smeared
out (also because of the smallness of the $L_I$-effects) that we can
take it that there is almost the same probability for finding the system
in state $| B_{t'} \rangle$ at any place in phase space allowed by
the conserved quantum numbers of the theory practically valid in the 
future era.
  \item[e)] Ignoring for simplicity the conserved quantities we
thus argued that with equal probability; equally distributed in phase
space, we have that $| B_{t'} \rangle$ will be any state.
  \item[f)] We can especially imagine that we have chosen a basis of
wave packet states $| w \rangle$ in the field configuration space so
that they fill smoothly the phase space
-- accessible without violating the conservation laws relevant --.
Taking these to be -- approximately -- orthonormal 
$\langle w| w' \rangle \approx \delta ww'$ we clearly get for the 
average expectation of the projection operator
\begin{eqnarray}\label{4.2}
P_{B_{t'}} = | B_{t'} \rangle \langle B_{t'} |
\end{eqnarray}
the estimate
\begin{eqnarray}\label{4.3}
\mathrm{av} (| B_{t'} \rangle \langle B_{t'} |)=
\frac{1}{N} \sum_{w} |w \rangle \langle w| \simeq \frac{1}{N} \underline{1}
\end{eqnarray}
where $N$ is the number of states in the basis
\begin{eqnarray}\label{4.4}
| w \rangle , ~w=1, \,2,\,\cdots , \,N.
\end{eqnarray}

That is to say we have argued for that our weak $S_I$-influence in future
combined with an assumed approximate ergodicity leads to that we can
approximate 
\begin{eqnarray}\label{4.5}
| B_{t'} \rangle \langle B_{t'} | \approx \frac{1}{N} \underline{1}
\end{eqnarray}
in practice for all $t'$ at least a bit later than the earliest
Big Bang.
\end{enumerate}

The crude estimate that we could replace $|B_t \rangle \langle B_t |$
by $\frac{1}{N}\underline{1}$ derived as formula
$(|B_t \rangle \langle B_t | \approx \frac{1}{N}\underline{1})$
were based on that $L_I$ were in practice small.

\section{Deriving a more usual probability formula}
We shall now make use of approximation (\ref{4.5}) for the ``future
factor" in the functional integral in order to obtain an expression
rewriting the formulas like (\ref{3}), (\ref{4}) and
(\ref{17}) and (\ref{18}) into expressions analogous to (\ref{19}).

The calculation is in fact rather trivial, starting say from the most
general of our postulated expressions (\ref{17}):
\begin{eqnarray}
\langle \CO (\varphi|_t , \,\Pi|_t )\rangle
&=& \frac{\int \e^{iS} \CO (\varphi|_t , \,\Pi|_t ) \CD \varphi
}{\int \e^{iS}\CD \varphi}
\nonumber\\
&=& \frac{\langle B_t | \CO (\varphi|_t , \,\Pi|_t ) | A_t \rangle
}{\langle B_t | A_t \rangle}
\nonumber\\
&\stackrel{\mathrm{trivial\, step}}{=}&
\frac{\langle A_t | B_t \rangle 
\langle B_t |\CO (\varphi|_t , \,\Pi|_t ) |A_t \rangle
}{\langle A_t | B_t \rangle \langle B_t | A_t \rangle}
\nonumber\\
&\stackrel{\mathrm{using\, (\ref{4.5})}}{=}&
\frac{\langle A_t |\frac{1}{N} \underline{1} \CO 
(\varphi|_t , \,\Pi|_t ) |A_t \rangle
}{\langle A_t |\frac{1}{N} \underline{1} |A_t \rangle}
\nonumber\\
&=& \frac{\langle A_t | \CO (\varphi|_t , \,\Pi|_t ) |A_t \rangle
}{\langle A_t | A_t \rangle}
\end{eqnarray}
which is the completely usual quantum mechanical expression
for the expectation value of the operator $\CO (\varphi|_t , \,\Pi|_t )$
in the wave functional state $|A_t \rangle$.

With this expression we see that we should be allowed, as we anyway would
expect, to use $|A_t \rangle$ as the quantum state of the universe.

It should be noted though that our $|A_t \rangle$ is
{\it in principle} calculable from the ``theory" when as we shall of course,
count also the $S_I$-expression as part of the theory.
In this way our model is widely more ambitious than usual quantum mechanics:

We have -- much like the Hartle-Hawking no boundary proposal --
a functional integral (\ref{6}) delivering in principle the wave functional
$|A_t \rangle$.
In usual quantum mechanics the wave function is left for the experimental
physicist to find out from his somewhat difficult job of preparing the state.
In practice we would presumably have to let him be so helped by observation
and arrangements under the preparation that we almost leave to him the usual
job.
We should, however, have in mind that in preparing a state one will usually
need to trust that some material is a rather pure chemical substance or
that no disturbing cosmic radiation spoils the preparation.
These kinds of trusts are usually based on some empirical experience which
in turn makes use of that big assembles of pure substances are 
easily/likely  available and that generally cosmic ray has low intensity.
Such trusts however, are at the very root connected with the starting
state -- the cosmology -- of our world.
But this starting state for practical purposes is in our model based on
the activity of our $L_I$ in early Big Bang times of the initial state
of the universe.

Thus it is even in the practical way of preparing a quantum state a lot of
reference to our $S_I$.

If, however, somehow the universe develops into states where $L_I$ is
no longer negligible we should expect corrections to such an 
approximation $(|B_t \rangle \langle B_t | \approx \frac{1}{N}\underline{1})$.

\section{Time development and $S_I$ corrections to $|B_t \rangle$}
{}From the definitions (\ref{6}) and (\ref{7}) of $|A_t \rangle$ and 
$|B_t \rangle$ it is trivial to derive the time development formulas
for these Hilbert space vectors (say for $t' > t$)
\begin{eqnarray}\label{6.1}
|A_{t'} \rangle &=& \int_{\mathrm{over\, time}-\infty \, \mathrm{to}\, t'}
 \e^{iS_{-\infty \,\mathrm{to}\, t'}} \CD \varphi
\nonumber\\
&=& \int_{\mathrm{over\,}t \, \mathrm{to}\, t'}
 \e^{iS_{t \,\mathrm{to}\, t'}} A_t [\varphi|_t] \CD \varphi
\nonumber\\
&=& \CU (t', \, t) |A_t \rangle
\end{eqnarray}
where $\CU (t', \, t)$ is the operator corresponding to the matrix
(with columns and rows marked by $\varphi|_t$ configurations)
\begin{eqnarray}\label{6.2}
\CU (\hat{\varphi}|^{'}_{t'},\, \hat{\varphi}|_t)
= \int_{\mathrm{over\,} t \, \mathrm{to}\, t'
\,\mathrm{with}\, \varphi|_{t'}=\hat{\varphi}|_{t'}
\,\mathrm{and}\, \varphi|_{t}=\hat{\varphi}|_{t}
}
\e^{iS_{t \,\mathrm{to}\, t'}}  \CD \varphi.
\end{eqnarray}
Similarly we have from (\ref{7}) for $t' > t$ again, first taking the 
complex conjugate of (\ref{7})
\begin{eqnarray}\label{6.3}
\langle \varphi |_t \,| B_t \rangle 
= \int_{\mathrm{beginning\, at}\, \varphi |_{t}}
\e^{-iS^{*}_{t \, \mathrm{to}\, +\infty}}\CD \varphi
\end{eqnarray}
and thus
\begin{eqnarray}\label{6.4}
\langle \varphi |_t \,| B_t \rangle 
= \int_{\mathrm{over}\, t \, \mathrm{to}\, t'}
\e^{-iS^{*}_{t \, \mathrm{to}\, +\infty}}
\langle \varphi |_{t'} | B_{t'} \rangle \CD \varphi
\end{eqnarray}
which can be written
\begin{eqnarray}\label{6.5}
|B_t \rangle = \CU_{\mathrm{with}\, L_I \to -L_I} (t', \,t)^+
\,|B_{t'} \rangle.
\end{eqnarray}
Here we used that e.g.
\begin{eqnarray}
S_{t \, \mathrm{to}\, +\infty} =
\int_{t}^{\infty} \mathrm{dt}
\int \mathrm{d}^3\vec{X} (\CL_R + i\CL_I)
\end{eqnarray}
where $\CL_R$ and $\CL_I$ are respectively the real and the imaginary
parts of the Lagrangian densities.
So
\begin{eqnarray}\label{6.6}
S^{*}_{t \, \mathrm{to}\, +\infty} =
\int_{t}^{\infty} \mathrm{dt}
\int \mathrm{d}^3\vec{x} (\CL_R - i\CL_I),
\end{eqnarray}
and now restricting ourselves for \{pedagogics/simplicity\} 
at first to boson fields we have (usually) that for them 
$\CL_R$ and $\CL_I$ are even order in the time derivatives which are
under latticification
\begin{eqnarray}\label{6.7}
\partial_t \varphi_{(t,\, \vec{x})}
\approx \frac{\varphi(t+\Delta t,\, \vec{x})- \varphi(t,\, \vec{x})
}{\Delta t} .
\end{eqnarray}
Thus conceived as operators between the configuration at the two close
by times $t$ and $t+ \Delta t$, i.e. with columns and rows marked by
$\varphi |_{t+\Delta t}$ and $\varphi |_t$ we have e.g.
\begin{eqnarray}\label{6.8}
(\CL_R + i\CL_I)^+ = \CL_R -i\CL_I
\end{eqnarray}
because
\begin{eqnarray}\label{6.9}
\CL_R^T = \CL_R ~~\mathrm{and}~~ \CL_I^T = \CL_I
\end{eqnarray}
and
\begin{eqnarray}\label{6.10}
\CL_R^* = \CL_R ~~\mathrm{and}~~ \CL_I^* = \CL_I.
\end{eqnarray}
In formula (\ref{6.5}) of course the meaning of the under symbol text
in the expression $\CU_{\mathrm{with}\,L_I \to -L_I}(t',\, t)^+$ is that
in addition to taking the Hermitian conjugation of $\CU(t',\, t)$ 
as defined by the matrix representation (\ref{6.2}) one shall shift the 
sign for all occurrences of the $L_I$-part of the Lagrangian or of the 
$L_I$-part of the Lagrangian density.
One should have in mind that it is easily seen that 
\begin{eqnarray}\label{6.11}
\CU(t',\, t)^{-1} =\CU_{\mathrm{with}\,L_I \to -L_I}(t',\, t)^+ .
\end{eqnarray}
Especially the ``usual" case of $L_I=0$ means that $\CU(t',\, t)$ becomes
unitary.
This relation (\ref{6.11}) together with (\ref{6.5}) and (\ref{6.1})
ensures that
\begin{eqnarray}\label{6.12}
\langle B_t | A_t \rangle =
\int \e^{iS_{-\infty \,\mathrm{to}\, +\infty}}\CD \varphi
\end{eqnarray}
can be true independent of the time $t$ chosen on the left hand side.

Since (6.1) represents a completely usual time development of the 
`wave function' $|A_t\rangle$ we have of course analogously to the usual 
theory
\begin{equation}
i\frac{d|A_t\rangle}{dt} = H|A_t\rangle 
\end {equation} 
where then H is the to the action   
\begin{equation}
S=S_R + S_I 
\end{equation}
corresponding Hamiltonian.
As we saw under point a) in section 2 formula (2.20) we can consider 
\begin{equation}
|A_t\rangle
\end{equation}
the wave function for the universe essentially. But really because of the 
normalizing denominator in (2.20) it is rather the normalized  $|A_t\rangle$, 
namely
\begin{equation}
|A_t\rangle_{norm} = |A_t\rangle/\sqrt{\langle A_t|A_t\rangle}
\end{equation}
which is the true wave function. 

It is important to remeark that precisely because we now find that we 
shall use the normalized wave function rather than $|A_t\rangle$ itself we do not 
get as could be feared a lack of conservation of probability due to the 
non-unitarity of the time development. Have in mind that the to a non-real 
action corresponding Hamiltonian H will not be Hermitean! But with the 
normalizaion comming from the $\langle A_t|A_t\rangle$ in the denominator in (2.20) the 
total probability will anyway remain unity.This result matches nicely with the 
from the slightly different start evaluated (9.22) below.

\vspace{.5 cm}

\noindent
{\huge \bf Part II, Second Trial of Interpretation}

\section{Second Interpretation of the functional integral}
Usually one only uses the functional integral over a time interval to
evaluate a transition matrix element from an initial time 
$t_i$ to a final time $t_f$
\begin{eqnarray}
U\left(\psi_f (\phi|_f),\psi_i (\phi|_i)   \right)
=\int  \CD^{\mathrm{ fixed\, time}} \phi|_f \int  \CD^{\mathrm{fixed\, time}} \phi|_i
\CD\phi\, \e^{iS_{t_{i}\,\mathrm{to}\,t_{f}}\left[\phi\right]}
\end{eqnarray}
where
\begin{eqnarray}
S_{t_{i}\,\mathrm{to}\,t_{f}}=
\int_{t_{i}}^{t_{f}} \int \CL(x)d^3\vec{x}dt
\end{eqnarray}
and the functional integral over $\CD\phi$ is restricted to $\phi$-functions
(field developments, or paths) which at times $t_i$ and $t_f$ respectively
coincides with $\phi|_i$ and $\phi|_f$ respectively.

In the present article we, however, have the ambition of having the
functional integral determine a priori not only the development with time 
but also say something about the initial conditions so that we a priori might 
ask for the probability of some dynamical variable $\CO$  say having certain
value $\CO$ at a certain time without imposing any initial conditions.
In order to obtain a formula or proceedure or how to obtain such
probabilities for what shall happen we have to assume such a formula.

We therefore need some intuitive and phenomenological guess leading to such
a formula/prescription.

In order to propose such a formula in a sensible way we shall first
consider a semiclassical approximation for our functional integral supposed
to be connected with and describing the development of the Universe
\begin{eqnarray}
\int \CD \phi \,\e^{i S[\phi]}
\end{eqnarray}
where we remember that in our model the $S[\phi]$ is not as usual real but 
is allowed to be complex.

\subsection{Semiclassical approach}
For first orientation let us imagine that the imaginary part of the action
$S[\phi]$ is effectively small in the sense that we can obtain the most
significant contributions to the functional integral by asking for saddle
points for the real part $S_R$.
That is we ask for field development solutions to the variational principle
\begin{eqnarray}
\delta S_R =0.
\end{eqnarray}
Without specifying the boundary conditions at $t \to \pm \infty$ in our
functional integral there should be (essentially) one solution for any
point in the (classical) phase space of the field theory described.
For the enumeration of the various development solutions $\phi$ we could
use the field and conjugate field configuration at any chosen moment of
time, to say.
However, now our hope and speculation is that the imaginary part should give
a probability weight distribution over the set ($\simeq$ phase space) of
these classical solutions.

\subsection{A first but wrong thinking}
It is clear that we must make a definition of an expectation value for 
function(al) $\CO$ say of the field development $\phi$ so that if a single
(semi) classical solution $\phi_{\mathrm{sol}}$ comes to be highly weighted then
this expectation value should be $\CO [\phi_{\mathrm{sol}}]$.

We might therefore at first think of 
\begin{eqnarray}\label{1.5}
\langle\CO\rangle = \frac{\int \CD \phi \CO [\phi] \,\e^{i S[\phi]}}
{\int \CD \phi \,\e^{i S[\phi]} }.
\end{eqnarray}

If really a single classical path contributed completely dominantly to both
numerator and denominator, then indeed we would obtain that this proposal
would obey
\begin{eqnarray}
\langle\CO\rangle = \CO[\phi_{\mathrm{sol\, dom}} ]
\end{eqnarray}
where $\phi_{\mathrm{sol\, dom}}$ is this single dominant solution.

It is however likely that if will be more realistic to imagine that 
there is a huge number of significant classical solutions $\phi_{\mathrm{sol}}$.
But then appears the ``problem" that in the expansion of the numerator 
functional integral into contributions from the various (semi) classical
solutions $\phi_{\mathrm{sol}\,i}$:
\begin{eqnarray}
\int \CD \phi \CO [\phi] \e^{i S[\phi]}
= \sum_{\phi_{\mathrm{sol}\,i\, \mathrm{all\, the\, classical\, solutions}}}
 \e^{i S[\phi_{\mathrm{sol}\,i}]} \CO[\phi_{\mathrm{sol}\,i}]
 \sqrt{\mathrm{det}_i}^{-1}
\end{eqnarray}
the various contributions contribute with quite different signs or rather
phases due to the appearance of the phase factor
$\e^{i {S_R} [\phi_{\mathrm{sol}\,i}]}$.
The proposal just put forward thus is not as it stands a usual average, 
it lacks the usual requirement of an average of being performed with 
a positive weight.
Rather the summation over the contribution becomes a summation with random
phases to a good approximation.
That means that if we classify in some ways the different solutions
$\phi_{\mathrm{sol}\,i}$ into classes, then what would sum up when such 
classes are combined would be the squared contributions rather than the 
contributions themselves.
In other words, if we define a contribution to
\begin{eqnarray}
\int \CD \phi \,\e^{i S[\phi]} \CO[\phi]
= \sum_{\phi_{\mathrm{sol}\,i}} \sqrt{\mathrm{det}_i}^{-1}
 \e^{i S[\phi_{\mathrm{sol}\,i}]} \CO[\phi_{\mathrm{sol}\,i}]
\end{eqnarray}

where

\begin{equation}
{\mathrm{det}_i} = \det \left ( 
\frac{\delta^2}{\delta \phi_1(x_1) \delta \phi_2(x_2)}
\right )
\end{equation}

from a certain class of semi classical solution $\CS_k$
then the quantities such as
\begin{eqnarray}
\int \CO \,\e^{i S} \CD\phi
\Big|_{\mathrm{from\, class\, \CS_k}}
\equiv 
 \sum_{\phi_{\mathrm{sol}\,i\in \CS_k}}
 \sqrt{\mathrm{det}_i}^{-1}
 \e^{i S[\phi_{\mathrm{sol}\,i}]} \CO[\phi_{\mathrm{sol}\,i}]
\end{eqnarray}
obey approximately
\begin{eqnarray}\label{asq}
&&\Bigg| \int \CO \,\e^{i S} \CD\phi
\Big|_{\mathrm{from\, class\, \CS_1}} \Bigg|^2 +
\Bigg| \int \CO \,\e^{i S} \CD\phi
\Big|_{\mathrm{from\, class\, \CS_2}} \Bigg|^2
\nonumber\\&&
\approx 
\Bigg| \int \CO \,\e^{i S} \CD\phi
\Big|_{\mathrm{from\, class\, \CS_{1}\cup \CS_{2}}} \Bigg|^2.
\end{eqnarray}
However we do not have a similar addition formula for numerical values as
\begin{eqnarray}
\Bigg| \int \CO \,\e^{i S} \CD\phi \Big|_{\mathrm{from\, class\, \CS_1}} \Bigg|^2,
\end{eqnarray} 
when they are not squared.
However, of course, we do have
\begin{eqnarray}
&& \int \CO \,\e^{i S} \CD\phi
\Big|_{\mathrm{from\, class\, \CS_1}}  +
 \int \CO \,\e^{i S} \CD\phi
\Big|_{\mathrm{from\, class\, \CS_2}} 
\nonumber\\&&
\approx 
 \int \CO \,\e^{i S} \CD\phi
\Big|_{\mathrm{from\, class\, \CS_{1}\cup \CS_{2}}} 
\end{eqnarray}
but this relation has terms of typically rather random phases.

\subsection{Approaching a probability assumption}
If we take $\CO$ to be a ``projection operator" in the sense of being a 
functional of $\phi$ only taking the values 0 and 1 then
$\int \CO \,\e^{i S} \CD\phi \Big|_{\mathrm{from\, class\, \CS_k}}$
should give the chance for solutions in the class $\CS_k$ to pass through the
configuration-class for which 
$\CO[\phi]=1$.
Because of the (random) phase and the lack of simple numerical additivity
mentioned if the foregoing subsection we are driven to assume that the probability
for $\phi$ being in the $\CO[\phi]=1$ region must be given by the squared 
contributions
\begin{eqnarray}
\Bigg| \int \CO \,\e^{i S} \CD\phi
\Big|_{\mathrm{from\, class\, \CS_k}} \Bigg|^2.
\end{eqnarray}

Calling the region in the space of $\phi$'s consisting of the $\phi$'s obeying
$\CO[\phi]=1$ with our ``project $\CO$", for region $M$, we get
\begin{eqnarray}
\mathrm{Prob}(M) \propto 
\Bigg| \int M \,\e^{i S} \CD\phi
\Big|_{\mathrm{from\, class\, \CS_k}} \Bigg|^2
\end{eqnarray}
for restriction to the class $\CS_k$.

This means using the probability of the complementary set $\bC M$ of $M$
\begin{eqnarray}
\mathrm{Prob}(\bC M) \propto 
\Bigg| \int \bC M \,\e^{i S} \CD\phi
\Big|_{\mathrm{from\, class\, \CS_k}} \Bigg|^2
\end{eqnarray}
and the additivity (\ref{asq})
\begin{eqnarray}
\Bigg| \int \,\e^{i S} \CD\phi
\Big|_{\mathrm{from\, class\, \CS_k}} \Bigg|^2 =
\Bigg| \int_M \,\e^{i} \CD\phi
\Big|_{\mathrm{from\, class\, \CS_k}} \Bigg|^2 +
\Bigg| \int_{\bC M} \,\e^{i S} \CD\phi
\Big|_{\mathrm{from\, class\, \CS_k}} \Bigg|^2 
\end{eqnarray}
we derive
\begin{eqnarray}
\displaystyle
\mathrm{Prob}(M)  &=&
\frac{\Bigg| \displaystyle\int_{M} \,\e^{i S} \CD\phi 
\Big|_{\mathrm{from\, class\, \CS_k}} \Bigg|^2}
{\Bigg|\displaystyle \int \,\e^{i S} \CD\phi
\Big|_{\mathrm{from\, class\, \CS_k}} \Bigg|^2
}.\nonumber\\
&=&
\frac{\Bigg|\displaystyle \sum_{\phi_{\mathrm{sol}\,i}\,\mathrm{in}\,M} \,\e^{i S} 
\sqrt{\mathrm{det}_i}^{-1}
\Big|_{\mathrm{from\, class\, \CS_k} }\Bigg|^2}
{\Bigg|\displaystyle \sum_{\phi_{\mathrm{sol}\,i}} \,\e^{i S} 
\sqrt{\mathrm{det}_i}^{-1}
\Big|_{\mathrm{from\, class\, \CS_k} }\Bigg|^2
}
\nonumber\\
&\stackrel{\mathrm{using\, random\, phases}}{\cong} &\displaystyle
\frac
{\displaystyle\sum_{\phi_{\mathrm{sol}\,i}\,\mathrm{in}\,M\cap \CS_k} \,\e^{-2 S_I} 
\mathrm{det}^{-1}
}
{\displaystyle
\sum_{\phi_{\mathrm{sol}\,i}\,\mathrm{in}\,\CS_k} \,\e^{-2 S_I} 
\mathrm{det}^{-1}
}.
\end{eqnarray}
Here in principle of a classical approximation the $\e^{-2 S_I}$ factor
is much more important than the ``quantum correction" $\mathrm{det}^{-1}$.
Thus we would ignore the determinant $\mathrm{det}^{-1}$ factor in first
approximation.

Then we arrived to the picture here that the probability distribution over
phase space - at some chosen time, that due to Liouville's theorem does not
matter - is given by $\e^{-2 S_I [\phi_\mathrm{sol}]}$ where $\phi_\mathrm{sol}$
is the classical field solution associated with the point in phase space for 
which $\e^{-2 S_I [\phi_\mathrm{sol}]}$ shall be the probability density.

\subsection{About the effect of $S_I$ in the classical approximation}
To appreciate the just given probability density 
$\e^{-2 S_I [\phi_\mathrm{sol}]}$ over phase space
\begin{eqnarray}
P\left(\phi |_{t_0}, \,\Pi |_{t_0} \right)
\CD \phi |_{t_0} \CD \Pi |_{t_0} 
\propto 
\e^{-2 S_I [\phi_\mathrm{sol}]} 
\CD \phi |_{t_0}, \CD \Pi |_{t_0}
\end{eqnarray}
one should have in mind that in the classical approximation of the universe
developing along a solution $\phi_\mathrm{sol}$ to the equations of motion
\begin{eqnarray}
\delta S_R = 0,
\end{eqnarray}
the development is given quite uniquely by the equations of motion.
The only place in which the imaginary part then comes in is in weighting with
various probability densities the various ``initial state data"
$\left(\phi |_{t_0}, \,\Pi |_{t_0} \right)$
-- i.e. the phase space point --.
Once you know the initial state of the (sub)system considered the equation
of motion determines everything in the classical approximation determined by
$S_R$ just described, the $S_I$ gets totally irrelevant.
In other words it is \underline{only} to know something about the
``initial state" that $S_I$ has relevance.
Here the usual terminology of ``initial state" shall especially in our model 
not be taken too seriously in as far as it with the word ``initial" refers
to a beginning moment, the Big Bang start say. No,
as we just mentioned one can use any moment of time $t_0$ for the 
description of the phase space describing the set of classical solutions
$\phi_{\mathrm{sol}}$.
This $t_0$ time does not have to be the first moment
-- even if such one should exist --.
Rather we can use any moment of time as $t_0$.
In the usual theory we would tend to use  $t_0$ being the initial moment and 
the state at this moment should then be one of very low entropy describing
our start of universe state.
However, in our model there is the rather unusual feature that the probability
weight $\e^{-2 S_I [\phi_{\mathrm{sol}}] }$ 
is given via a functional $S_I [\phi_{\mathrm{sol}}]$ 
depending on how the solution $\phi_{\mathrm{sol}}$ behaves at all
different times $t$ and not only at $t_0$.
Since we by the classical equations of motion can calculate the whole time
development $\phi_{\mathrm{sol}}$ from the fields and their conjugate
$\left(\phi |_{t_0}, \,\Pi |_{t_0} \right)$ at some chosen time $t_0$, 
we can of course also consider $\e^{-2 S_I [\phi_{\mathrm{sol}}] }$ 
as a function of only the data at $t_0$, 
$\left(\phi |_{t_0}, \,\Pi |_{t_0} \right)$.

So it is only by the fact that in our model
$\e^{-2 S_R [\phi_{\mathrm{sol}}] }$ is a rather \underline{simple} function
of $\phi_{\mathrm{sol}}$ and thus because of the often chaotic development 
of the fields by the classical equations of motion typically a complicated
function(al) of the time $t_0$ data.
With a more usual model one might think of the initial state in a 
``first moment" $t_0$ would be specified by some sort of cosmological model
or no boundary condition.
In this case the probability density should be rather simple in terms of the
first moment data.
The \underline{simplicity} of $\e^{-2 S_R [\phi_{\mathrm{sol}}] }$ as 
functional of the $\phi_{\mathrm{sol}}$-behavior even at late times to
some extend is extremely dangerous for our model showing observable effects
not observed experimentally.
Indeed an especially high probability for initial states leading to a 
special sort of happenings today say would look as a hand of God effect
seeking to arrange just this type of happenings to occur.
In practise we never know the state of the universe totally at a moment of
time.
So there would usually be possibilities to adjust a bit the initial 
conditions. That ciould then in our model have happened in such a way 
that events or things to happen in the future gets arranged, if it can be done 
so as to organize especially big $\e^{-2 S_R [\phi_{\mathrm{sol}}] }$ i.e.
an especially low $S_I$.
So a priori there would in our model be such ``hand of God effects".
In a later section we shall, however, invent or propose a possible
explanation that could make the era of today be of very little significance
for the value of $S_I$ so that in the first approximation it mainly the early
time features of a solution that counts for its probability density
\begin{eqnarray}
\e^{-2 S_I [\phi_{\mathrm{sol}}]}
\sim f(\phi_{\mathrm{sol}} |_{\hbox{``early times''}} )
= f(\mathrm{early\, time\, part\, of\, \phi_{\mathrm{sol}}})
\end{eqnarray}

\subsection{Relation to earlier publications}
We have earlier published articles working in the classical approximation
seeking to produce a model behind the second law of thermodynamics by
assigning a probability $P$ over the phase space of the Universe.
It were also there the point that this probability density $P$ in our
model should depend in the same way on the state at all times.
We already proposed that this $P$ were obtained by imposing an imaginary
part for the action $S_I$.
According to the above we clearly have
\begin{eqnarray}
P \propto \e^{-2 S_I}.
\end{eqnarray}

\section{Suggestion of the quantum formula}
We already suggested above that if $M$ denotes a subset of paths, 
e.g. those taking values in certain subset of $\phi |_t$-configuration
space in a moment of time $t$, then the probability for the true path being
in $M$ would be 
\begin{eqnarray}
\mathrm{Prob}(M)=
\frac{|\int_{M} \e^{i S} \CD \phi |^2}
{|\int \e^{i S} \CD \phi |^2}.
\end{eqnarray}
We imagine the paths to be described by the field $\phi$ as function over
$\bR^4$, the Minkowski space.
Thus we could use such an $M$ to describe e.g. the project of the possible
development $\phi$ to some subspace of configuration space $M_i$ for a 
series of moments $t_i, \,i=1, 2, \cdots, n$.
In fact then we would have
\begin{eqnarray}
M = \left\{ \phi \in \{\mathrm{paths} \}\Big| 
\phi |_{t_i} \in M_i \,\mathrm{for\, all}\, i \right\}.
\end{eqnarray}
It would in this case be natural to think of the functional integral
\begin{eqnarray}
\int_M \e^{i S} \CD \phi 
\end{eqnarray}
as a product of a series of functional integral associated with the various
time intervals in the series of times
$-\infty < t_1 < t_2 < \cdots < t_n < \infty$.
In fact let us define
\begin{eqnarray}
\CU_{t_i \,\mathrm{to}\, t_{i+1}}(\phi |_{t_{i+1}}, \phi |_{t_{i}})\equiv 
\int_{\mathrm{BOUNDARY}~\atop \phi|_{t_{i}} \,\mathrm{and}\,
\phi |_{t_{i+1}} \,\mathrm{kept}}
\e^{i S_{t_i \,\mathrm{to}\, t_{i+1} [\phi]}} \CD \phi
. 
\end{eqnarray}
Here 
\begin{eqnarray}
S_{t_i\,\mathrm{to}\,t_{i+1}}[\phi]
=\int^{t_{i+1}}_{t_i}\int\CL(x)
d^3\vec{x} dt
\end{eqnarray}
and remember that we here have the complex $d(x)$, 
\begin{eqnarray}
\CL(x)=\CL_R(x)+i\CL_I(x).
\end{eqnarray}
We then can write
\begin{eqnarray}
&&
\int_M\e^{iS}\CD\phi
\nonumber\\&&
=\int_i\CD^{(3)}\phi|_{t_i}
\CU_{t_n\,\mathrm{to}\,\infty}(\phi|_\infty,\phi|_{t_n})
\theta_{M_n}[\phi|_{t_n}]
\CU_{t_{n-1}\,\mathrm{to}\,t_n}
(\phi|_{t_n},\phi|_{t_{n-1}})
\theta_{M_{n-1}}[\phi|_{t_{n-1}}]
\cdots
\nonumber\\&&
~~~~~
\cdots
\theta_{M_1}[\phi|_{t_1}]
\CU
(\phi|_{t_1},\phi|_{t_{-\infty}})
\end{eqnarray}
where
$\theta_i[\phi|_{t_i}]$
is the function
\begin{eqnarray}
\theta_i[\phi|_{t_i}]=
\left\{
  \begin{array}{ll}
  1     &\mbox{for}~\phi|_i\in\CU_i    \\
  0     &\mbox{for}~\phi|_i\notin\CU_i      \\
  \end{array}
\right.
\end{eqnarray}
We can also write this expression in language of a genuine operator
product
\begin{eqnarray}
\int_M\e^{iS}\CD\phi=
\CU_{t_n\,\mathrm{to}\,\infty}\theta_n
\CU_{t_{n-1}\,
\mathrm{to}\,t_n}\theta_{n-1}\cdots
\theta_1\CU_{-\infty\,\mathrm{to}\,t_1}~.
\end{eqnarray}
where the $\theta_i$'s are now conceived of as projection operators on
the space of wave functionals characterized by being zero outside $M_i$,
\begin{eqnarray}
\theta_i\psi(\phi|_{t_i})=
\theta_i(\phi|_{t_i})\psi(\phi|_{t_i})
=\left\{
  \begin{array}{ll}
 0      & \mbox{for}~\phi|_{t_i}\notin M_i   \\
 \psi(\phi|_{t_i})      & \mbox{for}~\phi|_{t_i}\in M_i    \\
  \end{array}
\right.
\end{eqnarray}
Here $\psi$ is a possible/general wave functional for the state of the universe,
in the formula presented at the moment $t_i$.

In this operator formalism our probability formula takes the form
\begin{eqnarray}\label{fpmp64}
\mathrm{Prob}(M)&=&
\mathrm{Prob}(M_1,M_2,\cdots,M_n)
\nonumber\\
&=&
\frac{\left|
\CU_{t_n\,\mathrm{to}\,\infty}\theta_n
\CU_{t_{n-1}\,\mathrm{to}\,t_n}\theta_{n-1}
\cdots\theta_1\CU_{-\infty\,\mathrm{to}\,t_1}
\right|^2}{\left|
\CU_{t_n\,\mathrm{to}\,\infty}
\CU_{t_{n-1}\,\mathrm{to}\,t_n}
\cdots
\CU_{-\infty\,\mathrm{to}\,t_1}
\right|^2}
\end{eqnarray}
Since we are anyway in the process of arguing along to
just make an assumption about how to interpret in terms of 
probabilities for physical quantities of our complex action
functional integral, we might immediately see that it would be
suggestive to extend the validity of this formula for probabilities for 
field variables to also be valid for distributions in the conjugate
fields $\Pi|_{t_i}$ or in combinations,
\begin{eqnarray}\label{fpmp67}
&&
\mathrm{Prob}(\CO_1\in\tilde M_1,
\CO_2\in\tilde M_2,\cdots,
\CO_n\in\tilde M_n
)
\nonumber\\&&
=
\frac{
\left|
\CU_{t_n\,\mathrm{to}\,\infty}
P_{\CO_n\in\tilde M_n}
\CU_{t_{n-1}\,\mathrm{to}\,t_n}
P_{\CO_{n-1}\in\tilde M_{n-1}}
\cdots
P_{\CO_{1}\in\tilde M_{1}}
\CU_{-\infty\,\mathrm{to}\,t_1}
\right|^2}{
\left|
\CU_{t_n\,\mathrm{to}\,\infty}
\cdots
\CU_{-\infty\,\mathrm{to}\,t_1}
\right|^2}
\end{eqnarray}

Provided this proposal is not inconsistent to assume, we will assume
it because it would be quite reasonable to assume that the analogous 
formula to (\ref{fpmp64}) should be valid for any change for variables
between $\phi$ and $\Pi$ in the formulation of our functional integral.

\subsection{Consistency and no need for boundary conditions}
It should be kept in mind that we expect that due to the presence of the 
imaginary part $S_I$ in the action $S$ it is not needed to require any
boundary conditions at $t \to \pm \infty$ so that we basically can
remove as not relevant the $\phi |_\infty$ and $\phi |_{- \infty}$
boundaries which one would at first have considered to be needed in the 
expressions
$\CU_{-\infty\,\mathrm{to}\,t_1}(\phi|_{t_1},\phi|_{-\infty})$
and
$\CU_{t_n\,\mathrm{to}\,+\infty}(\phi|_{\infty},\phi|_{t_n})$.
The imaginary part $S_I$ is in fact expected to weight various contributions
so strongly different that whenever the by this weighting flavored
component in $\phi |_\infty$ say is at all allowed by a potential choice
of boundary condition then that contribution will dominate so much that all 
over contributions will be relatively negligible.
So after taking the ratio for normalization such as (\ref{fpmp67})
the choice of the boundary conditions for $\phi |_\infty$ and 
$\phi |_{- \infty}$ becomes irrelevant.
This irrelevance of the boundary conditions would indeed allow us to formally
put in according to our convenience of calculation whatever boundaries we
might like provided it does not precisely kill the boundary wave function
component flavored by the $S_I$.
For instance we could put in at the infinity density matrices taken to be
unity, since it does not matter anyway what we put and a unit matrix
$\rho_1 =1$ and $\rho_k =1$ would not suppress severely any state such as
the flavored one(s).

By this trick we could write our formula for probability
\begin{eqnarray}\label{2.13}
&&
\mathrm{Prob}(\CO_1\in\tilde M_1,
\CO_2\in\tilde M_2,\cdots,
\CO_n\in\tilde M_n
)
\nonumber\\&&
=\mathrm{Tr}(
\CU_{t_n\,\mathrm{to}\,\infty}
P_{\CO_n\in\tilde M_n}
\CU_{t_{n-1}\,\mathrm{to}\,t_n}
P_{\CO_{n-1}\in\tilde M_{n-1}}
\cdots
P_{\CO_{1}\in\tilde M_{1}}
\CU_{-\infty\,\mathrm{to}\,t_1}
\nonumber\\&&~~~~~~~~
\CU^\dag_{-\infty\,\mathrm{to}\,t_1}
P_{\CO_1\in\tilde M_1}
\cdots
P_{\CO_{n-1}\in\tilde M_{n-1}}
\CU^\dag_{t_{n-1}\,\mathrm{to}\,t_n}
P_{\CO_{n}\in\tilde M_{n}}
\CU^\dag_{t_{n}\,\mathrm{to}\,\infty}
)
\nonumber\\&&~~~
/\mathrm{Tr}
(
\CU_{t_n\,\mathrm{to}\,\infty}
\CU_{t_{n-1}\,\mathrm{to}\,t_n}
\cdots
\CU_{-\infty\,\mathrm{to}\,t_1}
\CU^\dag_{-\infty\,\mathrm{to}\,t_1}
\cdots
\CU^\dag_{t_{n-1}\,\mathrm{to}\,t_n}
\CU^\dag_{t_{n}\,\mathrm{to}\,\infty}
)~.~~~
\end{eqnarray}
Here the reader should have in mind that because of the imaginary part
in the action $S = S_R + i S_I$ the different development operators
\begin{eqnarray}
\CU_{t_i\,\mathrm{to}\, t_{i+1}}(\phi|_{t_{i+1}},\phi|_{t_{i}})
=\int_{\mathrm{WITH\,BOUNDARIES}~\atop
\phi|_{t_{i+1}}\,\mathrm{and}\,\phi|_{t_{i}}
\,\mathrm{at}\,t_{i+1}\,\mathrm{and}\,t_i\,
\mathrm{respectively} 
}\e^{iS[\phi]}\CD\phi
\end{eqnarray}
are in general not as usual unitary.
Therefore it is quite important to distinguish,
\begin{eqnarray}
\CU^\dag_{t_i\,\mathrm{to}\,t_{i+1}}
\stackrel{\mathrm{in\,general}}\neq
\CU^{-1}_{t_i\,\mathrm{to}\,t_{i+1}}.
\end{eqnarray}

\section{Practical Application Formulas}
\subsection{Practical application philosophy}\label{practical application}

Although in principle our theory is so much a theory of everything that
it should even tell what really happens and not only what is allowed by
the equations of motion, we must of course admit that even
we knew the parameters of both $S_I$ and $S_R$ it would be so exceedingly
hard to calculate what really happens that can\underline{not} do that.

We are thus first of all interested in using some reasonable approximations
to derive (in a spirit of a correspondence principle) some rules 
coinciding under practical conditions with the quantum mechanics
(or quantum field theory rather) rules we usually use.

Now as is to be explained in this section 
we can by means of requirements of monopoles and using the Standard Model
gauge symmetries and homogeneity of the Lagrangian in the fermion fields
argue that there will in the present era where Higgs particles are seldom
and Standard model applicable be only very small effects of $S_I$.
We also seem to have justified to make the same assumption for very huge time
spans in the future so that also until very far out in future the 
influence of $S_I$ is small.
Even if we imagine that in the very long run $S_I$ selects almost
uniquely the state or rather development - as we used above to argue that
the boundary conditions  $\phi |_{- \infty}$ and $\phi |_{+ \infty}$
were unimportant - then in the practical (i.e. rather near) future we
would expect the far future determination to deliver under an ergodicity
assumption an effective density matrix proportional to the unit matrix.
 
\subsection{Insertion of practical future treatment  into interpretation 
formula}
The above suggestion for the treatment of the practical future to be equally
likely in ``all" (practical) states is implemented by replacing what is
basically taking the place of a future density matrix
$\rho_f$ in our interpretation formula (\ref{2.13}) namely
\begin{eqnarray}
\rho_f \approx \CU^+_{t_{n}\,\mathrm{to}\,\infty}
\CU_{t_{n}\,\mathrm{to}\,\infty}
\end{eqnarray}
by a normalized unit density matrix
\begin{eqnarray}
\rho_f \approx \frac{1}{N}\underline{1},
\end{eqnarray}
where N is the dimension of the Hilbert space.
(In practice $N$ is infinite) So the interpretation formula becomes
\begin{eqnarray}\label{equation}
&&
\mathrm{Prob}(\CO_1\in\tilde M_1,
\CO_2\in\tilde M_2,\cdots,
\CO_n\in\tilde M_n
)
\nonumber\\&&
=\mathrm{Tr}(
P_{\CO_n\in\tilde M_n}
\CU_{t_{n-1}\,\mathrm{to}\,t_n}
P_{\CO_{n-1}\in\tilde M_{n-1}}
\cdots
P_{\CO_{1}\in\tilde M_{1}}
\CU_{-\infty\,\mathrm{to}\,t_1}
\nonumber\\&&~~~~~~~~
\CU^\dag_{-\infty\,\mathrm{to}\,t_1}
P_{\CO_1\in\tilde M_1}
\cdots
P_{\CO_{n-1}\in\tilde M_{n-1}}
\CU^\dag_{t_{n-1}\,\mathrm{to}\,t_n}
)
\nonumber\\&&~~~
/\mathrm{Tr}
(
\CU_{t_n\,\mathrm{to}\,\infty}
\cdots
\CU^\dag_{-\infty\,\mathrm{to}\,t_1}
\cdots
\CU^\dag_{t_{n-1}\,\mathrm{to}\,t_n}
)~.~~~
\end{eqnarray}
where we used that 
\begin{eqnarray}
P_{\CO_n\in\tilde M_n}
=P_{\CO_n\in\tilde M_n}^2~.
\end{eqnarray}

\subsection{Conditional Probability}
With formulas like (\ref{2.13}) or (\ref{equation}) we can easily form also 
conditional probabilities such as 
\begin{eqnarray}
&&
\mathrm{Prob}(\CO_{p+1}\in\tilde M_{p+1},
\cdots, \CO_n\in\tilde M_n
| \CO_1\in\tilde M_1 
,\cdots,
\CO_p\in\tilde M_p )
\nonumber\\&&
= \mathrm{Prob}(
\CO_1\in\tilde M_1
,\cdots,
\CO_p\in\tilde M_p
,\cdots,
\CO_n\in\tilde M_n)
\nonumber\\&&~~~
/\mathrm{Prob}(\CO_1\in\tilde M_1
,\cdots,
\CO_p\in\tilde M_p)
~.~~~
\end{eqnarray}
In order to determine what happens if we know the wave function in some 
moment.
Let us as an example consider the idealized situation of a case in which 
we know -- by preparation set up -- the whole state of the universe of
one moment of time.
This we could imagine being described by taking a series of 
$P_{\CO_i\in\tilde M_i}$ projection of the same moment of time,
the moment in which we suppose that we know the wave function.
For consistency and for being able to take the limit of them being at
same time -- and therefore with an ill-determined algebraic order in
(\ref{2.13}) we must assume these same time projectors to commute.
If we consider the situation that we already know that the system has
all these $\CO_i$ in the small regions.
$\tilde M_i$ because we know the wave function at their common time
$t_{\mathrm{com}}$ then we are after that discussing only the conditional
probabilities with the set of relations $\CO_i\in\tilde M_i$, 
$i=1, \cdots, p$, taken as fixed.

For simplicity let us consider the simple case that we just ask for 
if a variable $\CO_n$ at a later time being in $\tilde M_n$
or not then the conditional probability is in (\ref{2.13}) form
\begin{eqnarray}
&&
\mathrm{Prob}(\CO_n\in\tilde M_n
| \psi )
\nonumber\\&&
=
\mathrm{Tr}
( \CU^\dag_{t_n\,\mathrm{to}\,\infty}
P_{\CO_n\in\tilde M_n}
\CU_{t_\mathrm{com}\,\mathrm{to}\,t_n}
P_{\CO_1\in\tilde M_1}
P_{\CO_2\in\tilde M_2}
\cdots
P_{\CO_p\in\tilde M_p}
\CU_{-\infty\,\mathrm{to}\,t_\mathrm{com}}
\nonumber\\&&~~~~
\CU^\dag_{-\infty\,\mathrm{to}\,t_\mathrm{com}}
P_{\CO_p\in\tilde M_p}
\cdots
P_{\CO_2\in\tilde M_2}
P_{\CO_1\in\tilde M_1}
\CU^\dag_{t_\mathrm{com}\,\mathrm{to}\,t_n}
P_{\CO_n\in\tilde M_n}
\CU^\dag_{t_n\,\mathrm{to}\,\infty})
\\&&~~
/\mathrm{Tr}(
\CU_{t_\mathrm{com}\,\mathrm{to}\,t_n}
P_{\CO_1\in\tilde M_1}
\cdots
P_{\CO_p\in\tilde M_p}
\CU_{-\infty\,\mathrm{to}\,t_\mathrm{com}}
\CU^\dag_{-\infty\,\mathrm{to}\,t_\mathrm{com}}
P_{\CO_p\in\tilde M_p}
\cdots
P_{\CO_1\in\tilde M_1}
\CU^\dag_{t_\mathrm{com}\,\mathrm{to}\,t_n}~.
\nonumber
\end{eqnarray}
Herein we can substitute
\begin{eqnarray}
|\psi \rangle \langle \psi|= 
P_{\CO_1\in\tilde M_1}
P_{\CO_2\in\tilde M_2}
\cdots
P_{\CO_p\in\tilde M_p}
\end{eqnarray}
and obtain using as usual for traces Tr(AB)=Tr(BA) 
\begin{eqnarray}
&&
\mathrm{Prob}(\CO_n\in\tilde M_n
| \psi )
\nonumber\\&&
=
\langle \psi | 
\CU^\dag_{t_\mathrm{com}\,\mathrm{to}\,t_n}
P_{\CO_n\in\tilde M_n}
\CU^\dag_{t_n\,\mathrm{to}\,\infty}
\CU_{t_n\,\mathrm{to}\,\infty}
P_{\CO_n\in\tilde M_n}
\CU_{t_\mathrm{com}\,\mathrm{to}\,t_n}
| \psi \rangle
\langle \psi |
\CU_{-\infty\,\mathrm{to}\,t_\mathrm{com}}
\CU^\dag_{-\infty\,\mathrm{to}\,t_\mathrm{com}}
| \psi \rangle
\nonumber\\&&~~
/( \langle \psi |
\CU^\dag_{t_\mathrm{com}\,\mathrm{to}\,\infty}
\CU_{t_\mathrm{com}\,\mathrm{to}\,\infty}
| \psi \rangle 
\langle \psi |
\CU_{-\infty\,\mathrm{to}\,t_\mathrm{com}}
\CU^\dag_{-\infty\,\mathrm{to}\,t_\mathrm{com}}
| \psi \rangle )
\nonumber\\&&
\equiv  \langle \psi |
\CU_{t_\mathrm{com}\,\mathrm{to}\,t_n}
P_{\CO_n\in\tilde M_n}
\CU^\dag_{t_n\,\mathrm{to}\,\infty}
\CU_{t_n\,\mathrm{to}\,\infty}
P_{\CO_n\in\tilde M_n}
\CU_{t_\mathrm{com}\,\mathrm{to}\,t_n}
| \psi \rangle
/ \langle \psi |
\CU^\dag_{t_\mathrm{com}\,\mathrm{to}\,\infty}
\CU_{t_\mathrm{com}\,\mathrm{to}\,\infty}
| \psi \rangle~.
\nonumber\\
\end{eqnarray}
We may rewrite this expression in a suggestive way of the how it is modified
relative to usual quantum mechanics by defining the final state density
matrix from time $t_n$
\begin{eqnarray}
\rho_{f\,\mathrm{from}\,t_n}
\equiv 
\CU^\dag_{t_n\,\mathrm{to}\,\infty}
\CU_{t_n\,\mathrm{to}\,\infty}
\end{eqnarray}
in the following way 
\begin{eqnarray}
\mathrm{Prob}(\CO_n\in\tilde M_n
| \psi )=
\frac{ \langle \psi |
\CU_{t_{\mathrm{com}}\,\mathrm{to}\,t_{n}}
P_{\CO_n\in\tilde M_n}
\, \rho_{f\,\mathrm{from}\,t_{n}}\,
P_{\CO_n\in\tilde M_n}
\CU^\dag_{t_{\mathrm{com}}\,\mathrm{to}\,t_n }|\psi\rangle
}{
\langle \psi |
\CU_{t_\mathrm{com}\,\mathrm{to}\,t_n}
\, \rho_{f\,\mathrm{from}\,t_n}\,
\CU^\dag_{t_\mathrm{com}\,\mathrm{to}\,t_n}|\psi\rangle
}
\end{eqnarray}
Here $\CU_{t_\mathrm{com}\,\mathrm{to}\,t_n}$ is really a non unitary
$S$-matrix or development matrix for the time interval $t_\mathrm{com}$
at which we have $\psi$ to $t_n$ at which we look for $\CO_n$.
It is easy to see that we could have replaced $P_{\CO_n\in\tilde M_n}$
by a large set of commuting projections at a second common time $t_f$
destined to the single wave function $| \psi_f \rangle$ and thus
allowing to replace the series of projections put in place of
$P_{\CO_n\in\tilde M_n}$ by $| \psi_f \rangle \langle \psi_f |$.
Then we get for the probability of the transition from
$|\psi\rangle$ to $|\psi_f\rangle$
\begin{eqnarray}\label{pm99}
\mathrm{Prob}\left(|\psi_f\rangle\big||\psi\rangle\right)&=&
\frac{\langle\psi|\CU_{t_\mathrm{com}\,\mathrm{to}\,t_n}|\psi_f\rangle
\langle\psi_f|\rho_{f\,\mathrm{from}\,t_n}|\psi_f\rangle
\langle\psi_f|\CU^\dag_{t_\mathrm{com}\,\mathrm{to}\,t_n}|\psi\rangle
}{\langle\psi|\CU_{t_\mathrm{com}\,\mathrm{to}\,t_n}\rho_{f\,\mathrm{from}\,t_n}
\CU^\dag_{t_\mathrm{com}\,\mathrm{to}\,t_n}
|\psi\rangle}
\nonumber\\
&=&\left|
\langle\psi|\CU_{t_\mathrm{com}\,\mathrm{to}\,t_n}|\psi_f\rangle
\right|^2
\frac{\langle\psi_f|
\rho_{f\,\mathrm{from}\,t_n}|\psi_f\rangle}
{\langle\psi|\CU_{t_\mathrm{com}\,\mathrm{to}\,t_n}
\rho_{f\,\mathrm{from}\,t_n}
\CU^\dag_{t_\mathrm{com}\,\mathrm{to}\,t_n}|\psi\rangle
}
\end{eqnarray}
Now we can compare this expression with the usual transition probability
expression when $S$ is only real $=S_R$,
\begin{eqnarray}
\mathrm{Prob_\mathrm{usual}}
\left(|\psi_f\rangle\big||\psi\rangle\right)
&=&
\langle\psi|\CU_{t_\mathrm{com}\,\mathrm{to}\,t_n}|\psi_f\rangle
\cdot
\langle\psi_f|\CU^\dag_{t_\mathrm{com}\,\mathrm{to}\,t_n}|\psi\rangle
\nonumber\\&=&
\left|\langle\psi|\CU_{t_\mathrm{com}\,\mathrm{to}\,t_n}|\psi_f\rangle\right|^2
\end{eqnarray}
Denoting the transition operator
\begin{eqnarray}
S&\equiv&
\CU^\dag_{t_\mathrm{com}\,\mathrm{to}\,t_n}
\end{eqnarray}
this means we have
\begin{eqnarray}
\mathrm{Prob}
\left(|\psi_f\rangle\big||\psi\rangle\right)&=&
\left|\langle\psi_f|S|\psi\rangle\right|^2
\cdot
\frac{\langle\psi_f|\rho_{f\,\mathrm{from}\,t_n}|\psi_f\rangle}
{\langle\psi|\rho_{f\,\mathrm{from}\,t_n}|\psi\rangle}
\end{eqnarray}
compared to the usual expression
\begin{eqnarray}
\mathrm{Prob}_\mathrm{usual}
\left(|\psi_f\rangle\big||\psi\rangle\right)&=&
\left|\langle\psi_f|S|\psi\rangle\right|^2 .
\end{eqnarray}
The deviations are thus the following:
\begin{enumerate}
  \item With our imaginary part in $S$ there is no longer unitality, i.e.
\begin{eqnarray}
S^\dag \neq S^{-1}.
\end{eqnarray}
The transition $S$ is calculated by the Feynmann path integral with the
full
\begin{eqnarray}
S = S_R + iS_I .
\end{eqnarray}
  \item There is the extra wright factor
 $\langle \psi_f | \rho_{f\,\mathrm{from}\,t_n}$ describing the effect of
 the happenings and the $S_I$ in the future of the ``final measurement"
 $| \psi_f \rangle$.
  \item There is the only on the initial state $| \psi \rangle$ dependent
``normalization factor" in the denominator
\begin{eqnarray}
\langle\psi|S^\dag\rho_{f\,\mathrm{from}\,t_n}S|\psi\rangle
\end{eqnarray}
This denominator is indeed a normalization factor normalizing the total
probability for reaching a complete set -- an orthonormal basis --
of final states  $| \psi_{fk} \rangle$, $k=1, 2, \cdots$
which we for simplicity choose as eigenstate of 
$\rho_{f\,\mathrm{from}\,t_n}$
so that $\langle | \psi_{fk} \rho_{f\,\mathrm{from}\,t_n} | \psi_{fk} \rangle$
gets a diagonal matrix. 
Then namely
\begin{eqnarray}
&&
\sum_{k=1,2,\cdots}\mathrm{Prob}\left(|\psi_{f,k}\rangle\big||\psi\rangle\right)
\nonumber\\&&=
\frac{1}{\langle\psi|S^\dag \rho_{f\,\mathrm{from}\,t_n}S|\psi\rangle}\cdot
\sum_k\left|\langle\psi_{f,k}|S|\psi\rangle\right|^2
\langle\psi_{f,k}|\rho_{f\,\mathrm{from}\,t_n}|\psi_{f,k}\rangle
\nonumber\\&&=
\frac{1}{\langle\psi|S^\dag \rho_{f\,\mathrm{from}\,t_n}S|\psi\rangle}
\sum_k
\langle\psi|S^\dag|\psi_{f,k}\rangle
\langle\psi_{f,k}|\rho_{f\,\mathrm{from}\,t_n}|\psi_{f,k}\rangle
\langle\psi_{f,k}|S|\psi\rangle
\end{eqnarray}
which by using that the off diagonal elements of 
$\rho_{f\,\mathrm{from}\,t_n}$ were chosen to be zero can be rewritten as a 
double sum -- i.e. over both $k$ and $k'$ --
\begin{eqnarray}
&&
\sum_{k}\mathrm{Prob}\left(|\psi_{f,k}\rangle\big||\psi\rangle\right)
\langle\psi|S^\dag\rho_{f\,\mathrm{from}\,t_n}S|\psi\rangle
\nonumber\\&&=
\sum_{k,k'}
\langle\psi|S^\dag
|\psi_{f,k}\rangle\langle\psi_{f,k}|\rho_{f\,\mathrm{from}\,t_n}
|\psi_{f,k'}\rangle\langle\psi_{f,k'}|
S|\psi\rangle
\nonumber\\&&=
\langle\psi|S^\dag
\rho_{f\,\mathrm{from}\,t_n}
S|\psi\rangle
\end{eqnarray}
\end{enumerate}

Thus we see that this denominator just ensures that the total probability 
for all that can happen at time $t_n$ starting from $| \psi \rangle$
at $t_\mathrm{com}$ becomes just one.

\subsection{Simplifying formula for conditional probability by approximating
future}
We have already suggested that we should approximate
\begin{eqnarray}
\rho_{f\,\mathrm{from}\,t_n}\approx \frac{1}{N}\underline{1}
\end{eqnarray}
provided the future after $t_n$ is so that we can practically consider the
$S_I$-effects small or so much delayed into the extremely far future that
our above ergodicity argument can be used.
By such an approximation we remove the deviation number 2 above given by
$\langle  \psi_{f}| \rho_{f\,\mathrm{from}\,t_n} | \psi_{f} \rangle$
because we approximate this matrix element
$\langle  \psi_{f}| \rho_{f\,\mathrm{from}\,t_n} | \psi_{f} \rangle$
by a constant as a function of $|\psi_f \rangle$.
After this approximation we get
\begin{eqnarray}
\mathrm{Prob}\left(|\psi_{f}\rangle\big||\psi\rangle\right)
=
\frac{\left|\langle\psi_f|S|\psi\rangle\right|^2}
{\langle\psi|S^\dag S|\psi\rangle}
\end{eqnarray}
We should have in mind that $S^+S \neq 1$ in general since with
the imaginary part of the action the Hamiltonian will be non-hermitean
and $S$ non unitary.
Thus the usual $\left|\langle\psi_f|S|\psi\rangle\right|^2$
would by itself not deliver total probability for what comes out of 
$|\psi\rangle$ to be unity.
Only after the division by the normalization
$\langle\psi|S^\dag S|\psi\rangle$
would it become normalized to unity.

\section{Can we make an unsquared form?}
The formulas for the extraction of probabilities from our Feynmann path 
integral with imaginary part of action $S_I$ also were derived by 
considerations of statistical addition with essentially random phases
of various classical path.
But our crucial formula, say (\ref{2.13}), for probabilities is
seemingly surprisingly complicated in as far as each projection operator
occurs twice in the trace in the numerator.
Even the  simplest example of asking if some variable $\CO$ at time
$t$ falls into the range $\bar M$ gets the expression
\begin{eqnarray}\label{*}
&&
\mathrm{Prob}(\CO\in\bar M)
\nonumber\\&&
=\mathrm{Tr}(\CU_{t\,\mathrm{to}\,\infty}P_{\CO\in\bar M}\CU_{-\infty\,\mathrm{to}\,t}
\CU^\dag_{-\infty\,\mathrm{to}\,t}P_{\CO\in\bar M}\CU^\dag_{t\,\mathrm{to}\,\infty})
/
\mathrm{Tr}(\CU_{-\infty\,\mathrm{to}\,\infty}\CU^\dag_{-\infty\,\mathrm{to}\,\infty})
\nonumber\\
\end{eqnarray}
containing the projection operator 
$P_{\CO\in\bar M}$
twice as factor in the expression.
If we make the approximation of no 
$S_I$-effects
in the 
$t$ to $\infty$
time range by taking
\begin{eqnarray}
\rho_{f\,\mathrm{from}\,t_n}=
\CU^\dag_{t\,\mathrm{to}\,\infty}\CU_{t\,\mathrm{to}\,\infty}
\end{eqnarray}
and approximating it by being proportional to the unit matrix then,
however, the two projection operators come together and we could formally
replace their product by just one of them.
So in the case of the in this way approximated future we could write
\begin{eqnarray}
\mathrm{Prob}(\CO\in\bar M)&=&
\mathrm{Tr}(P_{\CO\in\bar M}
\CU_{-\infty\,\mathrm{to}\,t}
\CU^\dag_{-\infty\,\mathrm{to}\,t})/
\mathrm{Tr}(\CU_{-\infty\,\mathrm{to}\,t}
\CU^\dag_{-\infty\,\mathrm{to}\,t}).
\end{eqnarray}
If $\CO$ were a variable among the variables used as the path-description
in the Feynmann path integral the formula (\ref{*}) would by
functional integral be written
\begin{eqnarray}
\mathrm{Prob}(\CO\in\bar M)&=&
\frac{\left|
\int P_{\CO\in\bar M}\e^{iS}\CD\phi
\right|^2}{\left|
\int \e^{iS}\CD\phi
\right|^2}.
\end{eqnarray}
Strictly speaking these Feynmann integrals should be summed over all
end of time configurations, but with significant
$S_I$ presumably this summation would be dominated by a  few ``true"
initial and states at 
$\pm\infty$ and the summation would not be so important.

So strictly speaking we have
\begin{eqnarray}
\mathrm{Prob}(\CO\in\bar M)&=&
\frac{\sum_\mathrm{init,final}\left|
\int^\mathrm{final}_\mathrm{initial} P_{\CO\in\bar M}\e^{iS}\CD\phi
\right|^2}{\sum_\mathrm{init,final}\left|
\int \e^{iS}\CD\phi
\right|^2}.
\label{pm115}
\end{eqnarray}

\section{The Higgs width broadening}
As an example of application of the 
$S_I$-caused modification of the usual transition matrices we may consider
the decay of a particle 
-- which we for reasons to be explained below take to be the Weinberg-Salam
Higgs particle --
which has from $S_I$ induced an imaginary term in the mass (or energy).
Let us say take this term to have the effect of delivering a term being
a positive constant number multiplied by $-i$ in the Hamiltonian.
In the Schrodinger equation
\begin{eqnarray}
i\frac{d\psi}{dt}=H\psi
\end{eqnarray}
such a term will cause the wave function $\psi$ to decrease with time
so that it will decay exponentially with time $t$.
If the particle in addition decays ``normally" into decay products,
say $b\bar b$ as the Higgs particles
do the exponential decay rate will be the sum 
$\Gamma_{\mathrm{normal}}+\Gamma_{S_I}$
of the $S_I$-induced width
$\Gamma_{S_I}$ and the ``normal" decay width
$\Gamma_\mathrm{normal}$.
Let us for simplicity take as an approximation that the real part of
the mass is very large compared to both the ``normal" and the
$S_I$-induced widths so that we can work effectively non relativistically
with a resting Higgs particle.
We can let it be produced in a short moment of time which is short
compared to the inverse widths
$\frac{1}{\Gamma_{S_I}}$ and $\frac{1}{\Gamma_\mathrm{normal}}$
while still allowing the particle may be considered at rest approximately.

If we at first used the ``usual" formula
$|\langle\psi_f|S|\psi\rangle|^2$ for the decay process and calculate the
total probability for the particle to decay into anything one will find 
that this probability is only
$\frac{\Gamma_\mathrm{normal}}{\Gamma_\mathrm{normal}+\Gamma_{S_I}}$
because the average lifetime has been reduced by this factor, namely
from $\frac{1}{\Gamma_\mathrm{normal}}$ to 
$\frac{1}{\Gamma_\mathrm{normal}+\Gamma_{S_I}}$.
Since of course the usual particle with 
$\Gamma_{S_I}=0$ will decay into something with just probability unity, 
we thus need a normalization factor
$\langle\psi|S^\dag S|\psi\rangle$ to rescale the total probability to
be (again) unity in our imaginary action theory.

By Fourier transforming from time $t$ to energy the Higgs decay time
distribution we obtain in our model again a Breit-Wigner energy
distribution
\begin{eqnarray}
P(E)&=&
\frac{\Gamma_\mathrm{normal}+\Gamma_{S_I}}
{2\pi\left[(E-m_\mathrm{Higgs})^2
+\left(\frac{\Gamma_\mathrm{normal}+\Gamma_{S_I}}{2}\right)^2\right]}
\end{eqnarray}

If indeed we effectively should have such an $S_I$-induced imaginary part
in the mass of the Higgs, then the Higgs-width could be made bigger than
calculated in the usual width $\Gamma_\mathrm{normal}$.
This is an effect that might have been already seen in the LEP-collider
provided one has indeed seen some Higgses in this accelerator.
Indeed there has been found an excess of Higgs-like events with masses
slightly below the established lower bound for the Higgs mass of
114 GeV/c. 

\subsection{The effect of the 
$\langle\psi_f|\rho_{f\,\mathrm{from}\,t_n}|\psi_f\rangle$
suppression factor}
As an example of a (perhaps realistic) case of an effect of the factor
$\langle\psi_f|\rho_{f\,\mathrm{from}\,t_n}|\psi_f\rangle$
we could imagine that two particles are coming together organized to
hit head on
-- say in a relative $s$-wave -
able to potentially form two different resonances of which say one is a
Higgs which as above is assumed to have an imaginary term in its mass.
Now it is easy to see that the 
$\rho_{f\,\mathrm{from}\,t_n}$
(here $t_n$ is the moment the collision just formed one of the two
resonances thought upon as physical objects) will have
\begin{eqnarray}
\langle\psi_{f\,\mathrm{Higgs}}|\rho_{f\,\mathrm{from}\,t_n}|\psi_{f\,\mathrm{Higgs}}\rangle
<
\langle\psi_{f\,\mathrm{all}}|\rho_{f\,\mathrm{from}\,t_n}|\psi_{f\,\mathrm{all}}\rangle
\end{eqnarray}
where
$|\psi_{f\,\mathrm{Higgs}}\rangle$ and $|\psi_{f\,\mathrm{all}}\rangle$
represent respectively the two possible resonances the Higgs and the
alternative resonance.
Compared to the usual calculation of the transition to one of the resonances
-- essentially the square of the coupling constant --
the Higgs-resonance will occur with suppressed probability because of the
$\langle\psi|\rho_{f\,\mathrm{from}\,t_n}|\psi\rangle$-factor in the formula
(\ref{pm99}).
Really if the collision were safely organized that the collision occurs
because of $s$-wave impact preensured the total probability for one or the
other of the two resonances to be formed would be with properly normalized
probability 1 because of the
$\langle\psi|S^\dag S|\psi\rangle$
normalization factor.
However, the effect of
$\langle\psi_f|\rho_{f\,\mathrm{from}\,t_n}|\psi_f\rangle$
would be to increase the probability to form the alternative resonance while
decreasing the formation of the Higgs.

\section{Approaching a more beautiful formulation}
Taking the regions in which $\CO$ may lie or not $\bar M$ as infinitesimally
extended we would the formula for the probability density in the form
\begin{eqnarray}
\mathrm{Prob}(\CO=\CO_0)&=&
\frac{\sum_{i,f}\left|\int_{\mathrm{BOUNDARY:}i,f}\e^{iS[\phi]}P_{\CO\in\bar M}\CD\phi\right|^2}
{\sum_{i,f}\left|\int_{\mathrm{BOUNDARY:}i',f'}\e^{iS[\phi]\CD\phi}\right|^2}
\nonumber\\
&\propto &
\frac{\sum_{i,f}\left|\int_{\mathrm{BOUNDARY:}i,f}\e^{iS[\phi]}\delta(\CO-\CO_0)\CD\phi\right|^2}
{\sum_{i,f}\left|\int_{\mathrm{BOUNDARY:}i',f'}\e^{iS[\phi]}\CD\phi\right|^2}
\label{mpns1}
\end{eqnarray}

We may claim that this kind of formula the probability density for finding
$\CO$ taking a value infinitesimally close to $\CO_0$ is a bit unaestetic
because of having the projection operator $P_{\CO\in\bar M}$ or the 
equivalent $\delta(\CO-\CO_0)$ occurring \underline{twice}
while one might have said it would be simpler to have just one
$\delta(\CO-\CO_0)$ or $P_{\CO\in\bar M}$ factor in the expression.

We should now seek to reformulate our expression with these factors
occurring twice into a simpler one with only single occurrence of 
$P_{\CO\in\bar M}$ or $\delta(\CO-\CO_0)$.
To perform this hoped for derivation we first argue that for nonoverlapping
$\CO$-value regions $\bar M_1$ and $\bar M_2$
\begin{eqnarray}
&&
\sum_{i,f}\left(
\int_{\mathrm{BOUNDARY}:i,f}P_{\CO\in\bar M_1}\e^{iS[\phi]}\CD\phi
\right)^*
\cdot
\int_{\mathrm{BOUNDARY}:i,f}P_{\CO\in\bar M_2}\e^{iS[\phi]}\CD\phi
\nonumber\\&&
\approx 0~~~
\mbox{for}~~\bar M_1\cap\bar M_2=\emptyset ~.
\label{mpns4}
\end{eqnarray}
This is argued for by maintaining that giving $\CO$ a different value at
time $t$ very typically by ``butterfly effect"
-- Lyapunov exponent -- will cause very different states $f$ and $i$ at
$\mp\infty$ respectively.
If the two factors in (\ref{mpns4}) have very different final $f$ and
initial $i$ states dominate at the boundaries and even random phases
the total sum is indeed much smaller than what one would obtain if
$\bar M_1$ and $\bar M_2$ were taken to be the same region
$\bar M_1=\bar M_2=\bar M$.
If we now use the zero expression (\ref{mpns4}) by adding such terms
into the numerator and analogously in the denominator of (\ref{mpns1})
we can formulate this expression for the probability of $\CO$ being in
$\bar M$ into an expression involving a summation over the value or region
for $\CO$ in one of the occurrences
\begin{eqnarray}
\mathrm{Prob}(\CO=\CO_0^{(2)})
&=&
\Bigg(
\sum_{i,f}\int d\CO_0^{(1)}
\left(
\int_{\mathrm{BOUNDARY}: i,f}
\delta(\CO-\CO_0^{(1)})\e^{iS[\phi]}
\CD\phi
\right)^*
\nonumber\\&&~~~\cdot
\int_{\mathrm{BOUNDARY}: i,f}
\delta(\CO-\CO_0^{(2)})
\e^{iS[\phi]} \CD\phi 
\Bigg)
\\&&
\Big/
\left(\sum
_{i',f'}\left(\int_{\mathrm{BOUNDARY}: i',f'}
\e^{iS[\phi]}\CD\phi\right)^*\int_{\mathrm{BOUNDARY}: i',f'}
\e^{ iS[\phi]} \CD\phi
\right)~.
\nonumber
\end{eqnarray}

But now obviously we have
\begin{eqnarray}
\int d\CO_0^{(1)}\delta(\CO-\CO_0)=1
\end{eqnarray}
and thus we get
\begin{eqnarray}
\mathrm{Prob}(\CO-\CO_0^{(2)})&=&
\frac{\sum
_{i,f}\left(\int_{\mathrm{BOUNDARY}: i,f}
\e^{iS[\phi]}\CD\phi\right)^*
  \int_{\mathrm{BOUNDARY}: i,f}
\delta(\CO-\CO_0^{(2)})\e^{iS[\phi]}\CD\phi          }
{\sum
_{i',f'}\left(\int_{\mathrm{BOUNDARY}: i',f'}
\e^{iS[\phi]}\CD\phi\right)^*  \int_{\mathrm{BOUNDARY}: i',f'}
\e^{iS[\phi]}\CD\phi  }
\nonumber\\
\end{eqnarray}
In this expression we have achieved to have
$\delta(\CO-\CO_0^{(2)})$ only occurring once as factor.
We could therefore trivially extract from it an expression for the
average of the $\CO$-variable
\begin{eqnarray}
\langle\CO\rangle&=&
\int\CO_0^{(2)}\mathrm{Prob}(\CO-\CO_o^{(2)})d\CO_0^{(2)}
\nonumber\\&=&
\frac{\sum
_{i,f}\left(\int_{\mathrm{BOUNDARY}: i,f}
\e^{iS[\phi]}\CD\phi\right)^*
}
{\sum
_{i',f'}\left(\int_{\mathrm{BOUNDARY}: i',f'}
\e^{iS[\phi]}\CD\phi\right)^*}
\nonumber\\&&
\cdot
\frac{\int_{\mathrm{BOUNDARY}: i,f}
\CO\e^{iS[\phi]}\CD\phi}{
\int_{\mathrm{BOUNDARY}: i',f'}
\e^{iS[\phi]}\CD\phi}
\end{eqnarray}
If we could somehow remove the after all identical complex conjugate functional
integrals
\begin{eqnarray}
\left(\int_{\mathrm{BOUNDARY}: i, f}\e^{iS[\phi]}\CD\phi\right)^*
\label{mp138}
\end{eqnarray}
and
\begin{eqnarray}
\left(\int_{\mathrm{BOUNDARY}: i', f'}\e^{iS[\phi]}\CD\phi\right)^*
\label{mp138 prime}
\end{eqnarray}
only deviating by the dummy initial and final state designations respectively
$(i,f)$ and $(i',f')$, then we could achieve the simple expression (\ref{1.5}).
But in order to argue for such removal being possible we would have to 
speculate say that some -- we could say the true --
boundary condition combination for the functional integrals 
(\ref{mp138}, \ref{mp138 prime}) completely dominates.
This is actually not at all unrealistic since indeed the $S_I$ will tend to
very few paths dominate.
In such a case of dominance we would have a set of dominant
$(f,i)$.
Presumably to make the chance that there should be such dominance we
should allow ourselves to be satisfied with a linear combinations of
$i$-state and of $f$-states to dominate.
But now if indeed we could do that and call these linear combinations
$(f_\mathrm{dom},i_\mathrm{dom})$, then we could approximate
\begin{eqnarray}
\langle\CO\rangle\approx 
\frac{\int_{\mathrm{BOUNDARY}\,f_\mathrm{dom},i_\mathrm{dom}}\CO\e^{iS[\phi]}\CD\phi}
{\int_{\mathrm{BOUNDARY}\,f_\mathrm{dom},i_\mathrm{dom}}\e^{iS[\phi]}\CD\phi}.
\label{mp141}
\end{eqnarray}
Now we would like not to have the occurrence in this expression of the rather 
special states $(f_\mathrm{dom},i_\mathrm{dom})$.
However, these dominant boundary conditions are precisely the dominant boundary
conditions for the denominator integral, because it were really just the 
complex conjugate for the latter for which we looked for the dominant 
boundaries.

So if we let the boundaries free then at least the denominator should become
dominantly just as if we had used the boundaries
$(f_\mathrm{dom},i_\mathrm{dom})$.
It even seems that because of the smoothness and boundedness of the variable
$\CO$ as functional of $\phi$ the dominant boundaries $(i, f)$ should not
be much changed by inserting an extra factor $\CO$ so that also by letting
the boundaries free in the numerator functional integral
$\int\CO[\phi]\e^{iS[\phi]}\CD\phi$ would
not change much the dominant boundaries from those of the same integral with
the $\CO$-factor removed.
But the removal of this $\CO$ leads to the denominator functional integral,
for which we already saw that the dominating boundary behavior were
$(f_\mathrm{dom},i_\mathrm{dom})$.
Thus we have argued that we can rewrite (\ref{mp141}) into
\begin{eqnarray}
\langle\CO\rangle=
\frac{\int\CO\e^{iS[\phi]}\CD\phi}{\int\e^{iS[\phi]}\CD\phi}
\label{mp145}
\end{eqnarray}
where it is understood that the boundaries for
$t\to\pm\infty$ are ``free".
Then we suggested they would automatically go to be dominated by
$(f_\mathrm{dom},i_\mathrm{dom})$ thus fitting on to the formulas with
double occurrence of $\delta(\CO-\CO_0)$'s.

The argumentation that the factor $\CO$ does not matter for the dominant
behavior at $\pm \infty$ may sound almost contradictory to our assumption
using the ``butterfly effect" to derive the rapid variation of (\ref{4}) 
which meant that
an insertion of $\delta(\CO-\CO_0)$ would drastically change 
behavior, including that of the boundary.

It is, however, not totally unreasonable that a sharp function
$\delta(\CO-\CO_0)$ which is zero in most places could modify the 
boundary conditions, while a smooth one $\CO$, almost never zero would
not modify them.
Basically we hope indeed for that the $S_I$-caused weighting is so
severely restricting the set of significant paths, that it practically
means that a single path, ``the realized path" is selected.
In this case the insertion of the factor $\CO$ into the functional
integral would just multiply it by the value of $\CO$ on ``realized path".
If you however insert $\delta(\CO-\CO_0)$ and it as most likely the case
$\CO_0$ is not the value of $\CO$ on the realized path then we kill by the 
zero-value of $\delta(\CO-\CO_0)$ at the realized path would totally kill
the dominant contribution.
Then of course the possibility for a completely different path is opened
and the orthogonality used in (\ref{mpns4}) gets realistic.

As conclusion of the just delivered derivation of (\ref{mp145}) 
we see that the starting point in the beginning the articles is indeed
consistent under the suggested approximations with the forms derived 
from the semiclassical start.

\section{The monopole argument for suppressing the $S_I$ in the
Standard Model}
We have already above in section 3
argued that due to the material in the present era, and the future too,
being either massless or protected from decay by in practice conserved
quantum numbers and due to weakness of the interactions the contribution
to $S_I$ from these eras must be rather trivial.

It were also for the above argument important that the non-zero mass 
particles were non-relativistic in these eras.
That above argument may, however, not be sufficient for explaining that
no effect of our $L_I$ or $S_I$ would have been seen so far.
We have indeed had several high energy accelerators such as 
ISR (=intersecting storage ring at CERN) in which massive particles
-- such as protons --
have been brought to run for days with relativistic speeds.
That means that they would during this running in the storage rings say 
have had eigentime contributions significantly lower than the coordinate
 time or rather the time on earth.
This would presumably easily have given significant contributions to 
$S_I$ which going to the exponent would suppress
-- or priori perhaps enhance --
the probability of developments, solutions, to equations of motion,
leading to the running of such storage rings.
Since the protons have not already been made to run around dominantly
relativistically we should deduce that most likely the storage rings
would lead to increasing $S_I$ and thus lowering of the probability
weight.
Thus one would expect that the initial conditions should have been so 
adjusted as to prevent funding for this kind of accelerators, at least
for them running long time.
Contrary to Higgs producing accelerators which have so far not been able
to work on big scale (may be L.E.P. produced a few Higgses for a short time)
the accelerators with relativistic massive particles have seemingly
worked without especially bad luck.
In order to rescue our model it seems therefore needed to invent a crutch
for it of the type that there are actually no $L_I$-contributions
involving the particles for which the massive relativistically running
accelerators have been realized.
We have actually two mechanisms to offer which at the end can argue away
our $L_I$ or $S_I$ effects for all the hitherto humanly produced or found
particles, leaving the hopes for finding observable effects
-- bad luck for accelerators, mysterious broadening of resonance peaks --
to experiments involving the Higgs particle or particles outside the
Standard Model.
The point is indeed that we shall argue away the effects of $S_I$ for
gauge particles and for Fermions (coupled to them).

The suppression rules to be argue for are:
\begin{enumerate}
  \item[1)] Supposing the existence of monopoles we deduce that the
corresponding full gauge coupling constants must be real, basically as
a consequence of the Dirac relation.
  \item[2)] For fields which like the Fermion fields in renormalizable
theories occur homogeneously in the Lagrangian density
$\CL_R+i\CL_I$ this Lagrangian
density can be shown to be zero by inserting the equations of motion.
\end{enumerate}

\subsection{Spelling out the suppression rules}
Spelling out a bit the suppression rules let us for the monopole based 
argument remind the reader that although we consider a complex Lagrangian
density $\CL_R+i\CL_I$ for instance 
the electric and magnetic fields and the four potential $A_\mu(x)$
for electrodynamics are still real as usual.
Now if we have fundamental monopoles there must exist corresponding Dirac
strings which, however, as is well known must be unphysical.
The explicit flux in the Dirac string must
to have the Dirac string unobservable 
-- to be unphysical --
be compensated by an at the string  singular behavior of the four 
potential $A_\mu$
around the Dirac string.
The singular flux to compensate the extra flux in the Dirac string can,
however, only be real since the $A_\mu$-field is real and it is
given by a curve integral
$\oint A_\mu dx^\mu$
around the Dirac string.
Now as is well known the fluxes mentioned equal the monopole charges.
Thus the monopole charge $g$ must be real.
But then the Dirac relation
\begin{eqnarray}
eg=2\pi n,~~~n\in \bZ
\end{eqnarray}
tells that also the electric charge $e$ must be real.
Now, however, in the formalism with the electric charge absorbed into
the four potential
$\hat A_\mu=eA_\mu$
the coefficient on the
$F_{\mu\nu}^2$-term in the Lagrangian density is
$-\frac{1}{4e^2}$
so that the pure electromagnetic, kinetic, Lagrangian density
\begin{eqnarray}
\left(\CL_R+i\CL_L\right)\Big|_\mathrm{pure\,e.m.}=
-\frac{1}{4e^2}F^2_{\mu\nu}
\end{eqnarray}
becomes totally real.
I.e.
\begin{eqnarray}
\CL_L|_\mathrm{pure\,e.m.}=0.
\end{eqnarray}
We may skip or postpone a similar argument for non-abelian, Yang Mills,
theories to another paper, but really you may just think of some abelian
subgroup and make use of gauge invariance.

Concerning the rule 2) for homogeneously occurring fields, such as the
Fermion
fields in renormalizable theories the trick is to use the equations of motion.
For example the part of the Lagrangian density $\CL_R+i\CL_I$ involving
a Fermion field $\psi$ is of the form
$\CL_F=Z\cdot\bar\psi(i\Ds -m)\psi$
where $Z$ is a constant and 
$D_\mu$ the covariant derivative and of course
$\Ds=\gamma^\mu D_\mu$.
This Fermionic Lagrangian density is homogeneous of rank two in the Fermion 
field $\psi$.
The Euler-Lagrange equations, the equations of motion for the Fermion
fields are derived from functional differentiation w.r.t. the field $\psi$
\begin{eqnarray}
\frac{\delta S}{\delta\psi(x)}=0
\end{eqnarray}
and end up giving equations of motion of the form
\begin{eqnarray}
\frac{\partial\CL_F}{\partial\psi}=0
\end{eqnarray}
or
\begin{eqnarray}
\frac{\partial\CL_F}{\partial\bar\psi}=0
\end{eqnarray}
(really these forms are only trustable modulo total divergences but that
is enough) leading as is well known to
\begin{eqnarray}
\bar\psi(i\Ds-m)=0
\end{eqnarray}
or
\begin{eqnarray}
(i\Ds-m)\psi=0.
\end{eqnarray}
But now it is a general rule that a homogeneous expression,
$\CL_F$ say, can be recovered from its partial derivatives
\begin{eqnarray}
\sum\frac{\partial\CL_F}{\partial\psi}\psi+
\sum\bar\psi\frac{\partial\CL_F}{\partial\bar\psi}
=\mathrm{rank}\cdot \CL_F
\end{eqnarray}
where rank is in the present case 
rank$=2$.
Such a recovering for homogeneous Lagrangian densities, however,
means that the Lagrangian density
-- at least modulo total divergences --
can be expressed by the equation of motions, which are zero,
if obeyed.
But then at least in the classical approximation the Lagrangian density
is zero at least modulo total divergences.
This means that the total 
$S_R+iS_I$ contribution from the just discussed homogeneous terms 
end up zero.
Especially the imaginary part also ends up zero, although
its form does not have to be zero.
It is only insertion of equations of motion that makes it zero.

\section{Conclusion}
We have put up a formalism for a non-unitary model based on extending the
Lagrangian and thereby the action
to be complex by allowing complex coefficients in the Lagrangian density
$\CL_R+i\CL_I$.

We used two starting points for how to extract probabilities and expectation
values from the Feynmann path way integral in our ambitious model that 
shall even be able to tell what really happens rather than just the 
equations of motion.
The first were the interpretation that an operator $\CO(t)$ should have the
expectation value
\begin{eqnarray}
\langle\CO\rangle=
\frac{\int\CO(t)\e^{iS[\phi]}\CD\phi}
{\int\e^{iS[\phi]}\CD\phi}
\end{eqnarray}
but this expression is a bit dangerous in as far as it is a priori not
guaranteed to be real even though the quantity 
$\CO(t)$ is real.
The second approach would rather have a series of projections onto small
regions $\bar M_i$
for operator $\CO_i(t_i)$ denoted
$P_{\CO_i\in\bar M_i}$
inserted into the functional integral but then this integral
is numerically squared for any combination $(i, f)$ of boundary
behaviors at respectively $-\infty$ and $+\infty$ times.
That is to say that the insertions are to be performed into the integral
$\int\e^{iS[\phi]}\CD\phi$
so as to replace the latter by
$\int\prod_iP_{\CO_i\in\bar M_i} \e^{iS[\phi]}\CD\phi$
just as in the first approach, but then one forms the numerical square summed
over the initial $i$ and final $f$ behaviors
\begin{eqnarray}
\sum_{i,f}\left(
\int_{\mathrm{BOUNDARY}: i,f}\e^{iS[\phi]}\CD\phi
\right)^*
\int_{\mathrm{BOUNDARY}: i,f}\e^{iS[\phi]}\CD\phi
\label{product}.
\end{eqnarray}

The probability distribution is then obtained by inserting the projection
operators into $both$ factors in (\ref{product}) and then as normalization
divide the (\ref{product})  having these insertion with (\ref{product})
not having the insertions.

Under some suggestive assumptions we argued that the two approaches
approximately will agree with each other.
The most important formula derived is presumably the formula to
replace usual unitary $S$-matrix or $\CU$-matrix transition between
two moments in time in our model.
This formula turns out for transition an initial state
$|\psi\rangle$
to a final
$|\psi_f\rangle$
to be
\begin{eqnarray}
\mathrm{Prod}(|\psi_f\rangle,|\psi\rangle)=
\frac{|\langle\psi_f|S|\psi\rangle|^2
\langle\psi_f|\rho_{f\,\mathrm{from}\,t_f}|\psi_f\rangle}
{\langle\psi|S^\dag S|\psi\rangle}
\end{eqnarray}

We used that to derive the broadening in our model of the Higgs-width.

As an outlook we may mention some of the expectations of our model 
used in a more classical language in our earlier publications: If the Higgs 
-- especially freely running Higgses --- decrease significantly the probabilty
(7.21) then the initial state should be organized so that Higgs production be 
largely avoided. This would actually make the prediction that some how or the 
other an accident will happen and the LHC-accelerator will be prevented from 
comming to full energy and luminosoty.

\section*{Acknowledgments}

The authors acknowledge the Niels Bohr Institute (Copenhagen) and 
Yukawa Institute for Theoretical Physics for their hospitality 
extended to one of them each.
The work is supported by 
Grand-in-Aids for Scientific Research on Priority Areas, 
Number of Area 763 ``Dynamics of Strings and Fields", 
from the Ministry of Education of Culture, Sports, Science 
and Technology, Japan.

 \end{document}